\documentclass[conference]{IEEEtran}
\IEEEoverridecommandlockouts
\usepackage{tikz}
\usepackage{amsmath}
\usepackage{booktabs} 
\usepackage{multirow}
\usepackage{color,soul}
\usepackage{algorithm,float}
\usepackage{amssymb,amsfonts}
\usepackage[noend]{algpseudocode}
\usepackage{graphicx}
\usepackage{comment}
\usepackage{enumitem,kantlipsum}
\usepackage{textcase}
\usepackage{makecell}
\usepackage{url}
\usepackage{hyperref}
\usepackage{enumitem}
\usepackage{subfig}

\usepackage{breakurl}

\def\BibTeX{{\rm B\kern-.05em{\sc i\kern-.025em b}\kern-.08em
    T\kern-.1667em\lower.7ex\hbox{E}\kern-.125emX}}
\begin{document}

\title{Peeler: Profiling Kernel-Level Events to Detect Ransomware}

\author{\IEEEauthorblockN{Muhammad Ejaz Ahmed, Hyoungshick Kim, Seyit Camtepe, and Surya Nepal
\thanks{M. E. Ahmed, S. Camtepe, and S. Nepal are with Data61 CSIRO, Sydney, Australia (e-mail: \{ejaz.ahmed, seyit.camtepe, surya.nepal\}@data61.csiro.au).  H. Kim is with the College of Software, Sungkyunkwan University (SKKU), Suwon, Korea (e-mail: hyoung@skku.edu)}
}}

\maketitle

\begin{abstract}
Ransomware is a growing threat that typically operates by either encrypting a victim's files or locking a victim's computer until the victim pays a ransom. However, it is still challenging to detect such malware timely with existing traditional malware detection techniques. In this paper, we present a novel ransomware detection system, called ``Peeler'' (\textbf{\underline{P}}rofiling k\textbf{\underline{E}}rn\textbf{\underline{E}}l -\textbf{\underline{L}}evel \textbf{\underline{E}}vents to detect \textbf{\underline{R}}ansomware). Peeler deviates from signatures for individual ransomware samples and relies on common and generic characteristics of ransomware depicted at the kernel-level. Analyzing diverse ransomware families, we observed ransomware's inherent behavioral characteristics such as stealth operations performed before the attack, file I/O request patterns, process spawning, and correlations among kernel-level events. Based on those characteristics, we develop Peeler that continuously monitors a target system's kernel events and detects ransomware attacks on the system. Our experimental results show that Peeler achieves more than 99\% detection rate with 0.58\% false-positive rate against 43 distinct ransomware families, containing samples from both crypto and screen-locker types of ransomware. For crypto ransomware, Peeler detects them promptly after only one file is lost (within 115 milliseconds on average). Peeler utilizes around 4.9\% of CPU time with only 9.8 MB memory under the normal workload condition. Our analysis demonstrates that Peeler can efficiently detect diverse malware families by monitoring their kernel-level events.
\end{abstract}


\maketitle

\section{Introduction} \label{sec:introduction}
Ransomware is a class of malware that has significantly impacted enterprises and consumers. The goal of such type of malware is to obtain financial gain by holding a victim's system or resources as a hostage through encrypting the victim's files (called \texttt{crypto} ransomware) or locking the victim's desktop screen (termed as \texttt{screen-locker} ransomware). Recent statistics of ransomware shows that in the first three quarters of 2019, 151.9 million ransomware attacks were launched targeting enterprises and consumers~\cite{ransomwarestats,al2018ransomware}. The average payment to release files to the victims spiked to \$84,116 in the last quarter of 2019, more than double what it was in the previous quarter~\cite{ransomwarepayment}. In 2019, the U.S. alone was hit by an unprecedented barrage of ransomware attacks that impacted more than 966 government agencies, educational institutions, and healthcare providers at a potential cost of around \$7.5 billion~\cite{ransomwareimpact}. As such, ransomware represents one of the most visible threats to enterprises as well as users. Therefore, a large number of proposals have recently been proposed to fight against ransomware as follows: machine learning models (e.g.,~\cite{sivakorn2019countering, kharraz2017redemption, continella2016shieldfs, homayoun2017know, sgandurra2016automated}), use of decoy files (e.g.,~\cite{mehnaz2018rwguard,gomez2018r,continella2016shieldfs}), use of a key escrow mechanism (e.g.,~\cite{kolodenker2017paybreak}) and file I/O pattern profiling (e.g.,~\cite{morato2018ransomware,kharaz2016unveil,scaife2016cryptolock,huang2017flashguard, milajerdi2019poirot, sgandurra2016automated,zhao2019tee,continella2016shieldfs, al20170}). 



However, these techniques have at least one of the following limitations: 
1) \textit{Localised visibility:} existing approaches with limited localized visibility would struggle to detect certain type of malicious activities. For example, ransomware may first attempt to delete Windows backup files, disable anti-malware, or real-time monitoring services on the victim's machine to remain undetected before actually starting to encrypt user files. Techniques relying on file I/O request patterns, cryptographic primitives, or network traffic would fail to detect these activities making it challenging to recover from the attack. If suspicious activities are detected timely, ransomware can be detected without data loss. 
2) \textit{No guarantee on data loss:} most approaches do not provide any recovery or minimal data loss guarantees, i.e., late detection after several files have already been encrypted, or the computer has been locked~\cite{kharaz2016unveil,mehnaz2018rwguard,cyberpoint,scaife2016cryptolock,sgandurra2016automated}.
Scaife et al.~\cite{scaife2016cryptolock} proposed CryptoDrop which is based on the premise that ransomware aggressively encrypts user files. Their approach is able to detect a ransomware attack after a median of ten file losses. REDFISH~\cite{morato2018ransomware} detects the ransomware activity when ten files are lost, and the detection time took around 20 seconds. Similarly, RWGuard~\cite{mehnaz2018rwguard} took 8.87 seconds on average to detect all malicious processes spawned by ransomware.
3) \textit{Not flexible:} Most crypto ransomware detection approaches~\cite{mehnaz2018rwguard,morato2018ransomware,scaife2016cryptolock,cyberpoint} are built upon either file I/O patterns or cryptographic primitives, however, these techniques are not applicable to detect screen-locker ransomware requiring a separate module to detect them. For instance, in addition to detect crypto ransomware (based on file I/O request patterns), Kharraz et al.~\cite{kharaz2016unveil} developed computer vision-based machine learning module just to detect screen-locker ransomware. 
Similarly, PayBreak~\cite{kolodenker2017paybreak} can decrypt files only for the ransomware families that use system provided crypto functions. 
Such inflexibilities in existing approaches make it difficult to incorporate for different types of malware.

Moreover, to collect a diverse set of system events efficiently and consume them in a real-time is in itself a challenging task~\cite{cyberpoint}. For instance, Windows OS audit policies (\texttt{auditpol.exe}) allow users to enable several types of system events, but those are only a subset of all the telemetry available in Windows systems limiting its visibility~\cite{roberto_blog}. There are several types of data sources available, but they can not be easily enabled using an audit policy. Similarly, there are certain APIs (e.g., \textit{Win32 Event Tracing} API, \textit{System.Diagnostic.Eventing}, \textit{TraceEvent}) that allow us to log system events interactively, but they incur higher CPU and memory overheads~\cite{cyberpoint}. For an effective ransomware detection scheme, diverse set of data sources should be easily accessible for analysis with minimum overhead.


To address the shortcomings identified above, we develop a novel ransomware detection system, called ``Peeler'' (\textbf{\underline{P}}rofiling k\textbf{\underline{E}}rn\textbf{\underline{E}}l -\textbf{\underline{L}}evel \textbf{\underline{E}}vents to detect \textbf{\underline{R}}ansomware), which tracks ransomware activities from a ransomware binary execution on a victim's computer to the display of a \textit{ransom note} on the victim's screen. 
For a successful attack, ransomware first aim to remain undetected by anti-malware services on a victim's computer. To achieve stealthiness, ransomware may disable real-time monitoring, archive scanning, or even delete its own binary from the disk.  After that, they launch the actual attack -- encryption of user files or locking users' desktop screens, and finally, payment guidance activities, e.g., change background display picture containing ransom payment note. Peeler exploit several behavioural characteristics observed during the execution of a ransomware and develop a set of methodologies by leveraging rule-based, file I/O patterns-based matcher, and machine learning models.

Peeler provides comprehensive visibility via broad and deep monitoring of current and historical security configurations and events associated with sensitive operations. 
Unlike existing approaches that either rely on APIs or command line tools to collect system events, Peeler extract required system events from native layer of Windows OS using Windows Event Tracing (ETW) framework. 
As a result, Peeler performs significantly better compared to existing APIs and tools. 
Moreover, Peeler can intercept events from system-wide Windows OS providers (e.g., there are more than 1,100 providers in Windows 10) providing greater visibility. Finally, the generated system events can consumed in a real-time for early detection. Our key contributions are summarized below:

\begin{itemize} [leftmargin=*]
    \item \textit{Accurate:} We develop a set of methodologies based on comprehensive analysis of ransomware from binary execution on victim's computer to the display of a \textit{ransom note} at the victim's computer screen, and design a fast and highly accurate ransomware detection system, called Peeler. It relies on suspicious commands, file I/O event patterns, and two classification models with 13 system behavior features to achieve both fast detection and high detection accuracy. Overall, Peeler achieves 99.52\% detection accuracy with a false positive rate of only 0.58\% against 43 ransomware families -- the largest dataset so far in terms of its diversity. Moreover, Peeler achieves 100\% and 99.5\% detection accuracy against crypto and screen-locker ransomware, respectively. 
    
    
    
    \item \textit{Fast:} We measure the detection time of Peeler against both crypto and screen-locker ransomware families. Peeler took 115.3 milliseconds on average to detect crypto ransomware. Compared to the best existing crypto ransomware detection solution, Peeler is about five times faster. In addition, Peeler took 16.4 seconds on average to detect screen-locker ransomware while the execution time of screen-locker ransomware to lock the victim's desktop screen completely is 302.8 seconds on average, demonstrating that Peeler can also detect screen-locker ransomware at an early stage before locking a victim's system.
    
    \item \textit{Transferable:} We evaluate Peeler (without new training) against 86 samples from 26 new and unseen ransomware families, including crypto, screen-locker, and general malware samples. Peeler achieved more than 95\% detection accuracy in detecting new and unseen ransomware samples. Additionally, Peeler also detected 9 out of 16 samples from general malware.

\end{itemize}

\section{Key characteristics to detect ransomware} \label{sec:key insights for ransowmare detection}

Typically, a ransomware attack consists of three stages: perform stealth operations to remain undetected, launch the actual attack, and display \textit{ransom note} after a successful attack. 
In the first stage, all necessary actions required to perform a successful ransomware attack without being detected are executed. For instance, ransomware may attempt to change Windows OS configurations (e.g., disable real-time monitoring) to remain undetected by anti-malware services. In the second stage, the actual attack is launched, e.g., adversaries delete Windows OS shadow copies to prevent the victim from restoring the Windows OS to the previous version. In the final stage, a ransom note, along with guidance, is provided to the victim.

Peeler exploits the differences in system behavioral characteristics between ransomware and benign applications. We collected kernel-level events generated by diverse ransomware samples and benign applications and analyzed those events to gain insights into the unique and inherent system behaviors of ransomware. 
This section explores those characteristics in detail.




\begin{table}[h]
    \centering
    \caption{Set of actions required to perform attacks.} \label{tab:stealth}
    \resizebox{3.3in}{!}{
    \begin{tabular}{p{2.2cm}|p{8.8cm}}
    \toprule
         \textbf{Goal} & \textbf{Action}\\
         \midrule
         \multirow{7}{*}{Stealth} & Delete the ransomware executable from the disk. \\
         & Kill the task responsible for running malware. \\
         & Set boot entry elements to ignore errors if there is a failed boot or failed checkpoint. \\
         & Modify registry values to hide files and filename extensions. \\
         & Turn off User Account Control (UAC) to remain undetected.\\
         & Disable archive scanning, real-time monitoring by anti-malware services. \\
         \midrule
         \multirow{9}{*}{Attack} & Delete Windows OS shadow copies to prevent victim from restoring Windows OS.\\
         & Disable automatic repair feature in Windows OS. \\
         & Start executing malicious tasks, e.g., encryption of user files, with highest privileges.\\
         & Launch encoded PowerShell commands to change configurations. \\
         & Stop or bypass detection by a number of popular anti-malware services including Windows OS Defender. \\
         & Deny the specific rights to delete child folders.\\
         \midrule
         \multirow{4}{*}{Payment guidance} & Change Windows OS wallpaper informing the victim about attack. \\
         & Change background display picture containing ransom payment notes. \\
         & Open read-me document such as notepad containing all information about ransom payment. \\
         \bottomrule
    \end{tabular}
    }
\end{table}
\vspace{-3mm}
\subsection{Malicious commands} \label{ssec: suspicious commands} 
To maximize the impact of the encryption, ransomware performs malicious activities with the following three goals. 

\subsubsection{Stealthiness} 

Ransomware tries to remain undetected by anti-malware services or Windows OS defenders running on the victim's computer. For example, they may disable the following services: runtime monitoring, archive scanning, automatic startup repair (see Table~\ref{tab:stealth}). Moreover, they may delete ransomware executable from the disk, stop all anti-malware services, or turn off User Account Control (UAC). 

\subsubsection{Infeasibility of recovery}
Ransomware deletes the shadow copies and the system's backup/restore data automatically created by Windows OS. For example, \texttt{vssadmin} is a default Windows OS utility that controls volume shadow copies of user files on a given computer. These shadow copies are regularly used as a recovery point, and additionally, they can be leveraged to re-establish or return the file to a previous state if they are destroyed or lost due to some reasons. Adversary exploits \texttt{Vssadmin} utility, by executing the command \texttt{vssadmin.exe delete shadows /all /quiet}, to delete Windows OS shadow copies, making it impossible to restore the system back to its previous state. Of course, there are also other ways to delete shadow copies such as via \texttt{PowerShell}, \texttt{wmic}, etc.
The ransomware can leverage Windows OS program \texttt{net.exe} commands to stop or bypass detection by several popular antivirus software, in addition to defeating Windows OS Defender, e.g., \texttt{net.exe stop avpsus /y} stops Windows OS process Kaspersky Seamless Update Service which is used by Thanos ransomware family.
Moreover, ransomware sometimes tries to kill the processes related to specific programs, such as SQL server, to initiate the encryption of the user files on which these programs were operating. After encrypting user files, ransomware shows a ransom note and payment guidelines.

\subsubsection{Post-attack guidance on ransom payment} 

Finally, a ransom note is displayed along with a read-me document to help the victim pay the ransom and restore the system files. The ransomware writer typically adds a registry key to the autorun path to show the ransom note window to achieve this.

Table~\ref{tab:stealth} lists the set of actions performed by typical ransomware to achieve respective goals. The actual list of commands used in real-world ransomware are listed in Table~\ref{tab:stealth} is given in Appendix~\ref{appendix: malicious commands}. 
By intercepting such malicious commands, ransomware could be effectively detected. For instance, Hendler et al.~\cite{hendler2018detecting, hendler2020amsi} proposed deep learning approaches to detect malicious \texttt{PowerShell} commands. However, in practice, malicious commands could be launched not only by \texttt{PowerShell}, but also from Windows OS legitimate utilities, such as \texttt{vssadmin}, \texttt{wmic}, etc. We consider all malicious command sources rather than relying on malicious commands executed from a specific program.


\subsection{File I/O patterns} \label{ssec: file io patterns}

\begin{figure}[!ht]
  \centering
  \includegraphics[width=3.5in,clip]{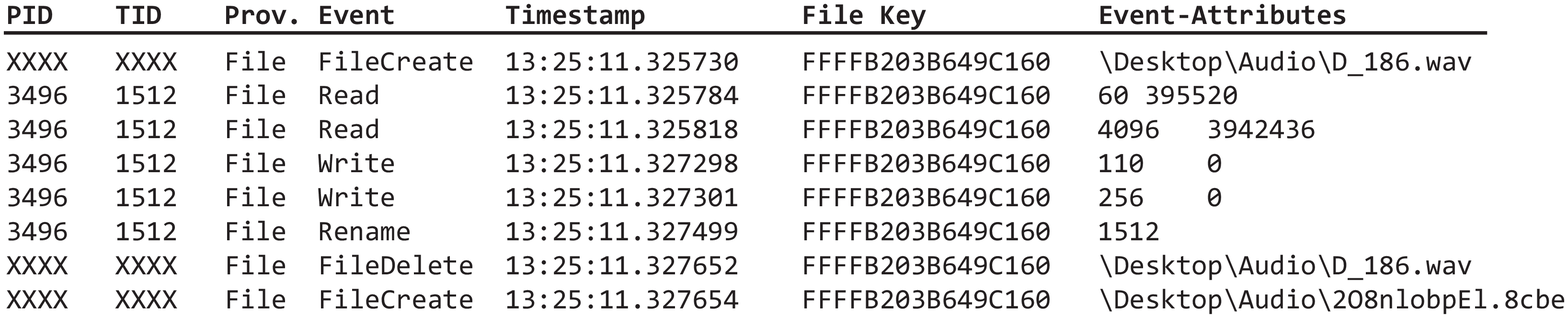}
  \caption{File I/O events generated by Cerber ransomware.}\label{fig:strat1A}
\end{figure}

Our analysis on collected events reveals that crypto ransomware samples typically encrypt a user's file by performing the following four steps: access the file (\textit{access}), read the content of the file (\textit{read}), write the encrypted content to a temporary memory or new file (\textit{write}), and overwrite/delete the user's original file (\textit{overwrite/delete}). For example, Figure~\ref{fig:strat1A} shows a sequence of file I/O events from a variant of Cerber. These events align with the observed four file I/O steps as follows: 1) in the \textit{access} step,  the ransomware sample accesses a file (\textit{D\_186.wav}) with the \texttt{FileCreate} event; 2) in the \textit{read} step, the ransomware sample reads the content of \textit{D\_186.wav} with the two \texttt{Read} events; 3) in the \textit{write} step, the ransomware sample writes the encrypted content to the same file with the two \texttt{Write} events; and 4) in the \textit{overwrite} step, the file is finally renamed with the \texttt{Rename}, \texttt{FileDelete}, and \texttt{FileCreate} events. As shown in Figure~\ref{fig:strat1A}, the \texttt{FileDelete} operation removes the content of the original file \textit{D\_186.wav} and \texttt{FileCreate} assigns a new name \textit{2O8nlobpEl.8cbe}. We note that the \texttt{File Key} remains the same for all events even though the original file's name and extension are changed.

We observe that most crypto ransomware samples follow similar steps to lock a victim's files. However, the locking strategy adopted by each ransomware sample can be different. For example, some families lock a file without creating a temporary file, whereas others choose to lock a file via a temporary file; some deletes the original file, whereas others decide to overwrite. We perform numerous trials and experiments and observe four \emph{generic} file I/O patterns that characterize behaviors of most crypto ransomware families. We briefly summarize our findings as follows.



\begin{figure}[!ht]
  \centering
  \includegraphics[width=3.4in,clip]{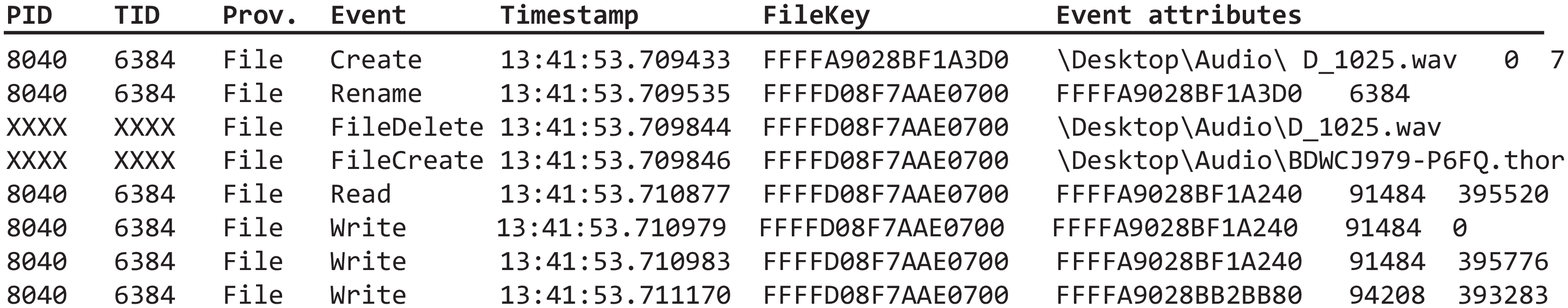}
  \caption{File I/O events generated by Locky ransomware.}\label{fig:strat1B}
\end{figure}

\begin{enumerate}[leftmargin=*]
    \item \textbf{Memory-to-File with Post-Overwrite:} As shown in Figure~\ref{fig:strat1A}, some crypto ransomware samples directly overwrite a user's file with its encrypted data without creating a new file. The I/O events pattern observed is to overwrite the encrypted data to the original file and then rename its file name. This pattern can be observed in the following ransomware families: Cerber, Keypass, Telsacrypt, and Gandcrab.
    
    \item \textbf{Memory-to-File with Pre-Overwrite:} This file I/O events pattern is similar to ``Memory-to-File with Post-Overwrite'' except that the original file is first renamed, and then the encrypted data is overwritten to that file. This pattern can be observed in samples from Locky ransomware family. Figure~\ref{fig:strat1B} shows a sequence of file I/O events generated from a sample in the Locky family. We can see that the original file is first renamed.
    
    \item \textbf{File-to-File with Delete:} Some crypto ransomware samples create a new file and copy the encrypted data of the original file to the new file instead of overwriting the original file itself. After completing the copy process, the original file is finally deleted. This pattern can be observed in the following families: InfinityCrypt, Dharma, Malevich, Sage, and Syrk. Figure~\ref{fig:strat2} in Appendix~\ref{appendix: file encryption patterns} shows a sequence of file I/O events generated from a sample in the InfinityCrypt family. We can see that the ransomware sample reads the file with the file key (\texttt{FFFFB203AFD146F0}) and writes it (in an encrypted form) to the file with the file key (\texttt{FFFFB203AFD14160}). After copying all the file content to the new file, the original file with the file key (\texttt{FFFFB203AFD146F0}) is deleted.
    
    \item \textbf{File-to-File with Rename and Delete:} This file I/O events pattern is similar to ``File-to-File with Delete'' except that the new file is renamed after copying the encrypted data of the original file to the new file. This pattern can be observed in samples from the WannaCry ransomware family. Figure~\ref{fig:strat3} in Appendix~\ref{appendix: file encryption patterns} shows a sequence of file I/O events generated from a sample in the WannaCry family. After copying all the file content to the new file, the file is renamed from \textit{nasa.txt.WNCRYT} to \textit{nasa.txt.WNCRY}.
\end{enumerate}

We show that most families from crypto ransomware follow one of the four file I/O patterns above described. 



\subsection{Application process tree} \label{ssec:application process tree}

\begin{figure}[!ht]
  \centering
  \includegraphics[width=3.4in,clip]{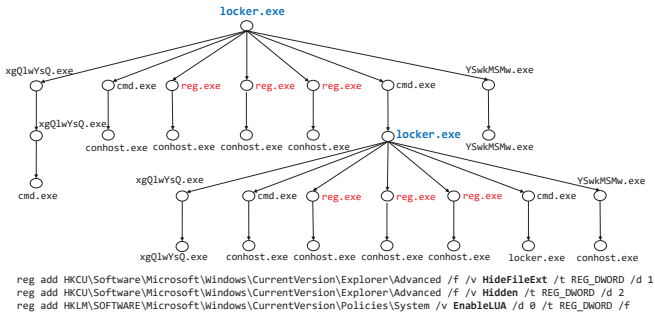}
  \caption{Application process tree of VirLock.}\label{fig:process tree ransomware}
\end{figure}

Applications can spawn one or more processes if it is needed. If a process in an application creates another process, then the creator process is called \textit{parent} process, and the created process is called \textit{child} process. We observe that some ransomware families spawns many child processes in a certain pattern compare to that of benign applications.
~For example, Figure~\ref{fig:process tree ransomware} shows a snapshot of the application process tree of VirLock ransomware during its execution. VirLock is self-reproducing ransomware that not only locks a victim's screen but also infects her files. Both behaviors -- self-reproducing and infecting files -- were observed in the application process tree of VirLock. We can see \textit{locker.exe} is replicated at level 3 of the tree.

To infect a victim's files, VirLock performs stealthy malicious activities to deceive victims. For instance, while creating files in the victim's computer, VirLock modifies the registry in the following ways: 1) disable Windows OS User Account Control (UAC), which is a feature that was designed to prevent unauthorized changes in desktop computers; 2) hide all files that are created on the victim's desktop; and 3) hide all created file extensions. As shown in Figure~\ref{fig:process tree ransomware}, three child processes (\textit{reg.exe} in red color) perform the actual registry modifications. The corresponding command line execution is shown at the bottom of Figure~\ref{fig:process tree ransomware}. 

\begin{figure}[!ht]
  \centering
  \includegraphics[width=3.4in,clip]{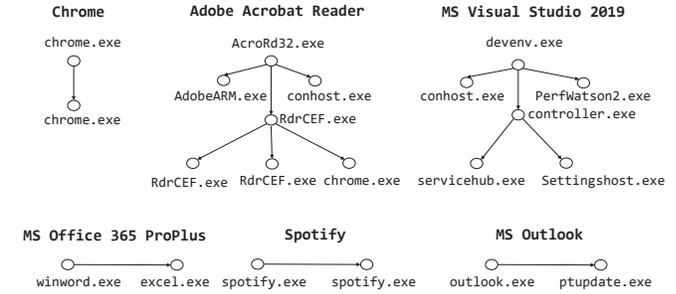}
  \caption{Application process trees of benign applications.}\label{fig:process tree benign}
\end{figure}

In contrast, most benign applications create far less number of spawned processes compare to the screen-locker ransomware. Figure~\ref{fig:process tree benign} shows the application process trees of six benign applications (\texttt{Chrome}, \texttt{Adobe Acrobat Reader}, \texttt{MS Visual Studio 2019}, \texttt{MS Office 365 ProPlus}, \texttt{Spotify}, and \texttt{MS Outlook}). 

Our observations on the behaviors of ransomware families show that more than 60\% of samples spawn multiple processes. For instance, VirLock ransomware sample spawns 44 processes on average.
On the other hand, our observations on more than 50 most popular benign Microsoft applications show that they spawn 16 processes on average, significantly smaller to that generated by a screen-locker ransomware sample.

\subsection{Correlations between system events} \label{ssec:system events correlation}

\begin{figure}[!h]
  \centering
  \includegraphics[width=3.4in,clip]{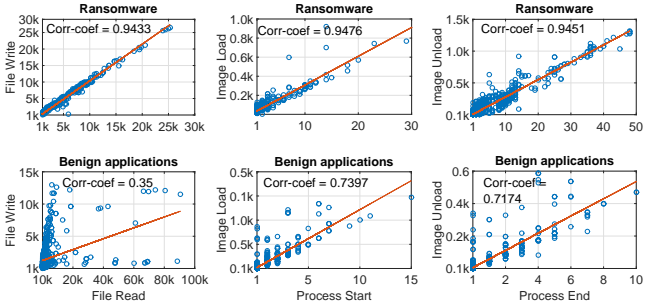}
  \caption{Correlations among some event types for ransomware and benign applications.}\label{fig:correlations}
\end{figure}

We analyzed the system events generated by ransomware samples and observed that there exist strong correlations between some events for the operations of ransomware.  For example,  all files read must be written (encrypted), which naturally shows a correlation between \texttt{Read} and \texttt{Write} events. Furthermore, similar correlations are exhibited among events collected from different providers such as \texttt{File}, \texttt{Process}, \texttt{Image}, and \texttt{Thread} because ransomware samples generate a large file \texttt{Read} and \texttt{Write} events to perform malicious tasks. Such relationships between certain events can be quantified by using the correlation coefficients of the events (see Table~\ref{tab:correlations}).
\begin{table}[h]
    \centering
    \caption{Correlation coefficients for some events.} \label{tab:correlations}
    \resizebox{3.0in}{!}{
    \begin{tabular}{lrr}
    \toprule
         \textbf{Events pair} & \textbf{Ransomware} & \textbf{Benign applications}  \\
         \midrule
         (File Read, File Write)& 0.9433 & 0.3500\\
         (Process End, Image Unload) & 0.9451  & 0.7174\\
         (Process Start, Image Load) & 0.9476  & 0.7397\\
         (Thread Start, Thread End) & 0.9560  & 0.6585\\
         \bottomrule
    \end{tabular}
    }
\end{table}
For example, a crypto ransomware sample generates \texttt{Read} and \texttt{Write} events regularly. As presented in Table~\ref{tab:correlations}, there exists a strong correlation between the number of \texttt{Read} events and the number of \texttt{Write} events. Such a correlation relationship may not appear in benign applications' \texttt{Read} and \texttt{Write} requests. Similarly, during ransomware execution, we observe the correlation between the number of \texttt{Start} processes and the number of image \texttt{Load} events, the correlation between the number of \texttt{End} processes and the number of image \texttt{Unload} events, and the correlation between the number of \texttt{Start} threads and the number of \texttt{End} thread events. These correlation coefficients are computed from the analysis performed on 206 ransomware samples and 50 most popular benign applications. Figure~\ref{fig:correlations} shows correlation among three pairs of events (\texttt{Read} and \texttt{Write}, \texttt{Start} and \texttt{Load}, \texttt{End} and \texttt{Unload}). We clearly observe strong correlations for ransomware compared to benign applications. Therefore, Peeler uses those correlations to detect ransomware. 

\section{System Design} \label{sec:system design}

Peeler is designed to minimize the overall damage by ransomware attacks using three different detection modules. Our design principle is to detect ransomware attacks as early as possible because the number of encrypted files (i.e., the victim's damage) can be increased if the attack detection is delayed. Therefore, our design principle is to detect it immediately by using a simple pattern matching first as soon as the first file is encrypted. We believe that the sacrifice of one file is our best effort scenario in the view of dynamic analysis because Peeler leverages file I/O event patterns that can be generated when a file is encrypted by a ransomware. 

To overcome the limitations of existing rule-based detectors, we consider not only specific rules to detect atomic observables (e.g., a known malicious file hash or a specific registry key modification) but also generic rules to identify ransomware's inherent behaviors. In practice, considering both specific and generic rules is essential in designing effective malware detection engines and the defenders' security posture~\cite{google_chronicle}. Based on this principle, Peeler's rules are categorized as follows: 1) key observables such as an indicator of compromise (e.g., malicious commands trying to delete Windows OS shadow copy); and 2) generic system event sequences (file I/O patterns discussed in Section~\ref{ssec: file io patterns}). 

To detect some sophisticated ransomware samples that are not detected by the rule-based approach, Peeler additionally adopts two different machine learning models by extracting behavioral features from process execution patterns (see Section~\ref{ssec:application process tree}) and correlations among various system events (see Section~\ref{ssec: feature extraction for system events}) for ransomware's activities. 


\subsection{Overview} \label{ssec:overview}



\begin{figure}[h]
  \centering
  \includegraphics[width=3.4in,clip]{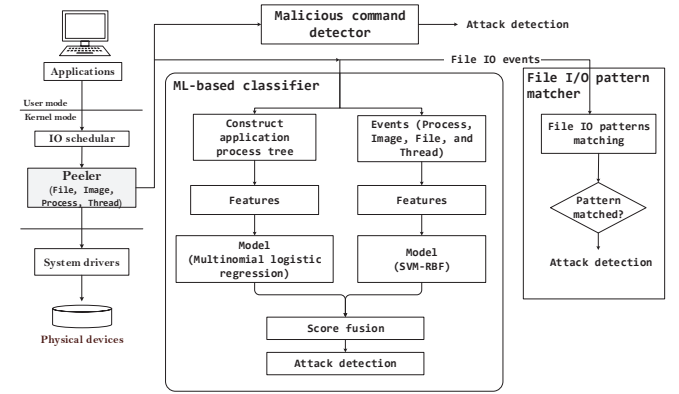}
  \caption{Overview of Peeler.}\vspace{-4mm}
\label{fig:system design}
\end{figure}

Peeler has three main ransomware detection modules: 1) malicious command detector, 2) file I/O pattern matcher, and 3) machine learning-based classifier. Figure~\ref{fig:system design} illustrates the overall design of Peeler. Peeler monitors system events continuously to detect ransomware attacks in real-time and uses them to perform ransomware detection. The malicious command detector uses pre-defined rules to check whether malicious commands are executed by processes in which the execution of those commands are mainly observed in ransomware activities. The file I/O pattern matcher takes \emph{kernel-level} file I/O events as input and looks for suspicious file I/O patterns that can be shown for ransomware. Once a pattern that characterizes ransomware behavior is detected, Peeler generates an alert. Machine learning-based classifier module extracts features from application process tree and system events providers (\texttt{Process}, \texttt{Image}, \texttt{File}, and \texttt{Thread}). The application process tree features are used to build a multinomial logistic regression model, whereas the system event features are used to build a Support Vector Machine (SVM) model. For detection, the scores from both classification models are fused as an ensemble approach, and then detection is performed.
 
Algorithm~\ref{algo:algo1} enlists all the key steps required for ransomware detection. A window size $W$ and a set of rules ($R_{s}$) are given as inputs to Algorithm~\ref{algo:algo1}. As shown in Figure~\ref{fig:system design}, the system events monitor collects kernel events and forwards them to three modules (the malicious command detector, the file I/O pattern matcher, and the machine learning-based classifier) in parallel. 
There are two differences between these modules: 1) malicious command detector and file I/O pattern matcher consumes events generated from the \texttt{Process} and \texttt{File} providers to identify malicious commands and suspicious file I/O patterns, respectively.
Whereas ML-based classifier utilizes events from all four providers (\texttt{File}, \texttt{Process}, \texttt{Image}, and \texttt{Thread}) and application process tree features to detect both crypto and screen-locker ransomware. 2) Malicious command detector and file I/O pattern matcher performs analysis on a per-event basis (Step 3 and 7, respectively), whereas ML-based classifier accumulates events for $W$ seconds (Step 11) and then processes them in a batch (Step 12). If all modules do not detect suspicious activities by ransomware, Peeler continuously monitors the system (Step 15).

\subsection{System events monitor} \label{ssec:monitoring agent}
Peeler provides a module called \emph{system events monitor} which relies on Event Tracing for Windows\footnote{ETW was first introduced in Windows 2000 and is now built-in to all Windows OS versions.} (ETW) framework. ETW is a built-in, general-purpose logging and diagnostic framework in Windows OS. It is \textit{efficient} (high speed and low overhead), \textit{flexible} (consume events in a real-time or log to a file), and provide greater \textit{visibility} into the system such that it allows to register to more than 1,100 subsystem providers~\cite{microsoftproviders} to receive and consume events. ETW framework is available for usage in both APIs and, command line tools and applications. The usage of ETW-based command line tools and applications are more common compared to ETW APIs (e.g.,\texttt{TraceEvent}, \texttt{System.Diagnostic.Eventing}, \texttt{Event Tracing}) to view system events~\cite{cyberpoint}. 

\begin{algorithm}[!ht]
\scriptsize
\caption{Overall Process of Peeler}\label{algo:algo1}
\begin{algorithmic}[1]
\Statex \textbf{Input: } $W$ and $R_s$
\While {true}
  \State System events monitor receives an $event$ from ETW. 
  \Statex /* \textbf{Input events to both detectors in parallel.}*/
  \If{event type is \texttt{Process Start}}
  \State Extract command from \texttt{commandLine} and match against rules in $R_s$.
  \If{command matched to a rule in $R_s$}
  \State Raise the alert and halt the process using \textit{PID}.
  \EndIf
  \EndIf
  \If{event is from \texttt{File} provider}
  \State \textit{CryptoMatcher} = FileIOPatternMatcher(\textit{event},\;$RE_{crypt}$)
  \If{\textit{CryptoMatcher}}  
  \State Raise the alert and halt the process using \textit{PID}.
  \EndIf
  \EndIf
  \State \textit{IncomingEvents} = Accumulate all events in a $W$ seconds window.
  \State \textit{MLClassifierLabel} = ML-basedClassifier(\textit{IncomingEvents})
  \If{\textit{MLClassifierLabel}}  
  \State Raise the alert and halt the process using \textit{PID}.
  \Else { Keep on monitoring the system. }
  \EndIf
\EndWhile
\end{algorithmic}
\end{algorithm}

Existing malware detection techniques either use built-in tools such as \texttt{Logman}, \texttt{TraceRpt}, \texttt{Event Viewer}, etc. or third party applications such as \texttt{Xperf}, \texttt{PerfView}, \texttt{Microsoft Message Analyser}, etc. 
However, the above mentioned tools suffer from at least one of the following issues: 1) they fail to parse events in real time, i.e., events are first logged to disk then parse, 2) they allow only subset of all telemetry available in Windows OS~\cite{roberto_blog}, and 3) high overhead in terms of CPU and memory. Table~\ref{tab:etwtools} compares existing ETW-based approaches for event collection with Peeler's \textit{system events monitor} in terms of performance metrics.
\begin{table}[h]
    \centering
    \caption{Tools and applications using ETW framework.}
    \resizebox{3.4in}{!}{
    \begin{tabular}{lccc} 
    \toprule
         \textbf{Metric} & \textbf{Command line tools and applications} & \textbf{ETW APIs} & \textbf{Peeler} \\
         \midrule
         Visibility & limited & limited &  greater\\
         Real-time & $\times$ & \checkmark & \checkmark \\
         Lightweight &\checkmark & $\times$ & \checkmark \\
         Native & $\times$ & $\times$ & \checkmark \\
         Static support & \checkmark & \checkmark & \checkmark \\
         \bottomrule
    \end{tabular}
    }
\label{tab:etwtools}
\end{table}

We designed a module called \textit{system events monitor} based on ETW. Our module is light-weight because it directly interacts with the native layer and performs filtering of the system events. In terms of visibility, Peeler extract events from the following providers: \texttt{Process, Image, File, Thread}. A unified data model (UDM) containing a set of events is extracted and then consumed to detect ransomware. The data obtained by Peeler's system events monitor are in the form of a continuous sequence of events $t_i$. An event is represented as:
$t_i = <PID, TID, Prov., EType, E_{timestamp}, E_{attrs}>,$ where PID is a process identifier, TID is a thread identifier corresponding to the process $PID$.
Prov. is provider name, EType is event name,$E_{timestamp}$ is the time of event occurrence, and $E_{attrs}$ is a set of attributes of the event $E_{name}$.

Each event has a schema describing the type of data contained in the event payload. The overall data schema is listed in Table~\ref{tab:systemevents}. 
To implement the system events monitor in Peeler, we used an open-source project \textit{krabsetw}~\cite{krabsetw}, which is a C++ library that simplifies interactions with ETW. We modified \textit{krabsetw} both at API and native layer to collect the events only needed for Peeler. 

\begin{table}[h]
    \centering
    \caption{Providers and events used in Peeler.}
    \resizebox{3.4in}{!}{
    \begin{tabular}{llcc} 
    \toprule
         \multirow{2}{*}{\textbf{Provider}} & \multirow{2}{*}{\textbf{Event}} & \multicolumn{2}{c}{\textbf{Event schema}} \\ \cmidrule{3-4}
         & & \textbf{Common attributes} & \textbf{Provider-specific event attributes}  \\
         \midrule
         \multirow{2}{*}{Process} & \multirow{2}{*}{Start, End} & \multirow{2}{*}{PID, TID, Prov., Event, Timestamp} & \multicolumn{1}{l}{SessionId, ParentId}\\ & & &\multicolumn{1}{l}{ImageFileName, CommandLine}\\ \midrule
         \multirow{3}{*}{File} & Read, Write & PID, TID, Prov., Event, Timestamp & \multicolumn{1}{l}{FileKey, FileObject, IoSize}\\ 
         & Rename, Delete & PID, TID, Prov., Event, Timestamp & \multicolumn{1}{l}{FileKey, FileObject}\\
         & FileCreate, FileDelete& PID, TID, Prov., Event, Timestamp & \multicolumn{1}{l}{FileObject, FileName} \\ \midrule
         Thread & Start, End & PID, TID, Prov., Event, Timestamp & \multicolumn{1}{l}{ParentId}\\ \midrule
         Image & Load, Unload & PID, TID, Prov., Event, Timestamp & \multicolumn{1}{l}{ImageSize, FileName} \\ 
         \bottomrule
    \end{tabular}
    }
\label{tab:systemevents}
\end{table}



\subsection{Malicious commands detector} \label{ssec:malicious command detector}
Peeler uses a component called \emph{malicious command detecter} to filter suspicious activities conducted by ransomware using the database for malicious commands which are needed to perform ransomware's activities (see Table~\ref{tab:stealth}). Peeler can collect malicious commands from the \texttt{Process}'s \texttt{Command line} argument (see Table~\ref{tab:systemevents}). Typically, the \texttt{commandLine} attribute of the \texttt{Process} contains the actual commands (see example commands in Table~\ref{tab:maliciouscommands} in Appendix~\ref{appendix: malicious commands}). For example, Windows OS utility \texttt{net.exe} can start, stop, pause or restart any service using the command \texttt{net.exe stop ServiceName} launched via convenient script/batch file or command prompt. If an adversary leverages such commands to stop several anti-malware services, Peeler can detect the process using pre-defined rules. Similarly, utility \textit{taskkill.exe} ends one or more tasks or processes. Typically, ransomware specify the image name of the process to be terminated (e.g., \textit{taskkill.exe /IM ImageName}). The \texttt{sc.exe} utility modifies the value of a service's entries in the registry and service control manager database. Ransomware may attempt to disable several defending services on the victim's machine before starting encryption. Peeler uses rules based on utility names and actions performed in order to detect ransomware infection.


\subsection{File I/O pattern matcher} \label{ssec:fileio pattern detector}



To effectively detect suspicious file I/O patterns, Peeler relies on detecting a sequence of suspicious file I/O events, as discussed in Section~\ref{ssec: file io patterns}. 
Algorithm~\ref{algo:algo2} is composed of three main stages for detecting crypto ransomware. 
The input to the Algorithm is events from file I/O provider. An event is specified as \textit{event} = $<$\textit{PID}, \textit{EType}, \textit{FileObject}, \textit{FileName}, \textit{FileKey}$>$ and with a set of regular expressions, $RE_{crypt}$. \textit{PID} and \textit{EType} refer to the process identifier and event type, respectively. \textit{FileObject} is a unique key assigned to every file, whereas \textit{FileName} contains the name of a file with full path. 
\textit{FileKey} is an identifier that is used to find the file on which an event is performed. For example, the event schema for \texttt{Read} and \texttt{Write} (see Table~\ref{tab:systemevents}) does not contain \textit{FileName} attribute. 
In order to obtain the file name, the \textit{FileKey} attribute of the \texttt{Read}/\texttt{Write} event is matched with the \textit{FileObject} attribute of the \texttt{FileCreate} event.


\subsubsection{Preparing the data}
The continuous stream of incoming file I/O events are first processed such that they are added to a separate list based on the file they are operating on. Whenever a user's file is accessed (i.e., \texttt{FileCreate} event is received), an events list \textit{FileEventsList} is created and the corresponding event (\texttt{FileCreate}) with respective attributes is added to that list. All subsequent file I/O operation events (such as \texttt{Read}, \texttt{Write}, \texttt{Rename}, \texttt{Delete}, \texttt{FileCreate}, and \texttt{FileDelete}) on that file are added to its \textit{FileEventsList} (Step 1 -- 6 in Algorithm~\ref{fig:strat2}).

However, it is challenging to accurately add all file I/O events corresponding to a user's file to its respective \textit{FileEventsList}. Because a single file may have multiple \textit{FileObject} keys and vice versa. For example, as shown in Figure~\ref{fig:strat2}, the \textit{FileObject} keys for \texttt{Read} and \texttt{Write} events on the file \texttt{D\_186.wav} are not the same, but the actual file on which the operations are performed is the same. To ensure that all events associated with a file are fully recorded, Peeler leverages both \textit{FileObject} and \textit{FileName} attributes to accurately identify a user's file.


\begin{algorithm}[!ht]
\scriptsize
\caption{FileIOPatternMatcher.}\label{algo:algo2}
\begin{algorithmic}[1]
\Statex \textbf{Input: } \textit{event} = $<$\textit{PID}, \textit{EType}, \textit{FileObject}, \textit{FileName}, \textit{FileKey}$>$
\Statex \textbf{Output: } \textit{Benign} and \textit{Ransomware}
\Statex
\Statex \textbf{Stage 1: Preparing the data}
\If{EType is `FileCreate'}
\If{FileObject and FileName are newly observed}
\State Create events list \textit{FileEventsList} for the newly observed file. 
\State Add \textit{event} to \textit{FileEventsList}.
\EndIf
\Else { add \textit{event} to respective file's \textit{FileEventsList}.}
\EndIf
\Statex
\Statex \textbf{Stage 2: File I/O patterns detection}
\If{the number of unique ETypes in the \textit{FileEventsList} is equal or greater than four}
\State Extract Etypes sequence from \textit{FileEventsList}. 
\If {Etypes sequence matched to at-least one of the file I/O patterns in Section~\ref{ssec: file io patterns}}
\State Flag the process PID in \textit{event}.
\State Ransomware = True
\EndIf
\Else{ continue}
\EndIf
\Statex
\Statex \textbf{Stage 3: Filtering false positives.}
\If{\textit{FilePath} are different in \textit{FileEventsList}'s events} 
\State Benign = True
\EndIf
\If{PIDs in \textit{FileEventsList}'s events are not same} 
\State Benign = True
\EndIf
\end{algorithmic}
\end{algorithm}

\subsubsection{File I/O pattern matcher}  

Suspicious file I/O patterns are detected by analyzing the sequence of events in the list \textit{FileEventsList} corresponding to a file. If the number of unique \textit{ETypes} is equal to or more than four (Step 6 in Algorithm~\ref{fig:strat2}), the events in the list are analyzed for suspicious patterns.  Peeler enforces the \textit{four unique events} constraint to reduce unnecessary computational overhead because at least four unique event types are required to encrypt a user file (see Figures~\ref{fig:strat1A}--\ref{fig:strat3}). 
If the sequence of incoming events in \textit{FileEventsList} matches with file I/O patterns in Section\ref{ssec: file io patterns}, i.e., file access, read, write (encrypted content), and file renaming operations are performed in order on a user's file, Peeler captures this sequence as the evidence of crypto ransomware activity for a security warning.

We note that the \textbf{Memory-to-File} operations can be distinguished from the \textbf{File-to-File} operations by using the \textit{FileObject} keys. When the same file is overwritten, we observe that the \textit{FileObject} key remains the same for all file I/O events, e.g., the sequence of events to encrypt the file \textit{D\_186.wav} are shown in Figure~\ref{fig:strat1A} for Cerber ransomware. For \textbf{File-to-File} operations, multiple files associated with \textit{FileObject} keys are needed. That is, in this case, the file I/O events from multiple files are accumulated into the same \textit{FileEventsList} list.

\subsubsection{Filtering false positives}
\label{sec: Filtering false positives}

We observe that some benign applications may generate ransomware-like file I/O patterns. For example, some benign applications may overwrite Windows OS's Activation Tokens file (\textit{tokens.dat}), which can lead to false positives because it generates file I/O events that look similar to \textbf{Memory-to-File} patterns. We applied the following two heuristics to reduce the possibility of such false-positive cases (see Stage 3 in Algorithm~\ref{algo:algo2}):

\begin{itemize}[leftmargin=*]
    \item If \textit{FilePath}s in \textit{FileEventsList} are not directing to the same file, this is ignored because the file encrypted must be in the same location.
    \item All the file I/O events must be performed by the same process (i.e., the same \textit{PID}) with exception for the \textit{system} and \textit{explorer.exe} processes. For example, event lists in Figures~\ref{fig:strat2} and~\ref{fig:strat3} show that \textit{system} process (\textit{PID} = $4$) is involved. Similarly, \textit{explorer.exe} is a system process that assists other processes in performing tasks.
\end{itemize}


\subsection{Machine learning-based classifier} \label{ssec:feature extraction}

Peeler uses two different machine learning-based (ML-based) classifiers with application process tree features, and system event features to detect ransomware samples that cannot be detected by the file I/O pattern matcher. Algorithm~\ref{algo:algo3} enlists all the steps for ML-based classifiers. The input to the algorithm is a set of events accumulated over $W$ seconds window. In our current implementation, we empirically set $W=5$ to optimize the speed-accuracy tradeoff. Features are extracted to build two machine learning models (Step 1--4 in Algorithm~\ref{algo:algo3}). The built machine learning models are then used to detect ransomware attacks (Step 5--7 in Algorithm~\ref{algo:algo3}). 

\begin{algorithm}[!ht]
\scriptsize
\caption{ML-basedClassifier.}\label{algo:algo3}
\begin{algorithmic}[1]
\Statex \textbf{Input: } \textit{IncomingEvents}
\Statex \textbf{Output: } \textit{Benign} and \textit{Ransomware}
\Statex 
\Statex \textbf{Stage 1: Feature sets extraction and model building.}
\State $FV_{MLR}$ = extract application process tree features from \textit{IncomingEvents} (see Table~\ref{tab:feature extraction}).
\State $FV_{SVM}$ = extract system events features from \textit{IncomingEvents} (see Table~\ref{tab:feature extraction}).
\State \textit{M1} = Build ML Multinomial logistic regression model with $FV_{MLR}$ feature set.
\State \textit{M2} = Build ML SVM-RBF model with $FV_{SVM}$ feature set.
\Statex 
\Statex \textbf{Stage 2: Attack detection.}
\State Extract feature set vectors for test samples.
\State Test $FV_{MLR}$ and $FV_{SVM}$ features on models \textit{M1} and \textit{M2}, respectively.
\State Fuse scores from models and provide the class label (either \texttt{benign} or \texttt{ransomware}) as output.
\end{algorithmic}
\end{algorithm}

\subsubsection{Building the first classifier with the application process tree features} \label{ssec: feature extraction for application process tree}



Windows OS applications are typically descended from the parent process called \textit{explorer.exe}. Therefore, all applications' processes share \textit{explorer.exe} as their parent process. As discussed in Section~\ref{ssec:application process tree}, we observed that several ransomware samples spawn  significantly larger number of child processes compared to benign applications. Based on this observation, Peeler constructs a machine learning model using the application process tree features. Peeler specifically extracts the following features from the application process tree: the number of processes, the number of unique processes, the number of threads created by processes, the maximum depth level of the tree, and the number of leaf nodes in the tree. For the first classifier using the feature set $FV_{MLR}$ (see Table~\ref{tab:feature extraction}), we selected a multinomial logistic regression (MLR) model because we do not assume linear relationships among the features in $FV_{MLR}$~\cite{bayaga2010multinomial} and MSR produces the best performance with those features.




\subsubsection{Building the second classifier with the system event features} \label{ssec: feature extraction for system events}

As discussed in Section~\ref{ssec:system events correlation}, Peeler leverages four providers' (\texttt{File}, \texttt{Process}, \texttt{Image}, and \texttt{Thread}) events exhibiting casualties. In total, the following four pairs of events are used: (\texttt{Read}, \texttt{Write}), (\texttt{Start}, \texttt{Load}), (\texttt{End}, \texttt{Unload}), and (\texttt{Start}, \texttt{End}). To capture these casualties simply, Peeler extracts frequency features (see Table~\ref{tab:feature extraction}) and train an SVM model based on feature set $FV_{SVM}$ for classification. We selected SVM with RBF kernel because it is lightweight and produces the best accuracy results with $FV_{SVM}$. 



\begin{table}[h]
    \centering
    \caption{Feature extraction.} \label{tab:feature extraction}
    \resizebox{2.4in}{!}{
    \begin{tabular}{l|l|l}
    \toprule
         \textbf{Feature set} & \textbf{Feature} & \textbf{Model}\\
         \midrule
         \multirow{5}{*}{$FV_{\text{MLR}}$} & \# of processes & \multirow{5}{*}{MLR}\\
         & Sum of threads from processes & \\
         & Maximum depth level of process tree & \\
         & \# of leaf nodes & \\
         & \# of unique process names & \\
         \midrule
         \multirow{8}{*}{$FV_{\text{SVM}}$} & \# of process start & \multirow{8}{*}{SVM-RBF} \\
         & \# of process end &\\
         & \# of DLL image loads & \\
         & \# of DLL image unloads &\\
         & \# of file reads & \\
         & \# of file writes & \\
         & \# of threads start & \\
         & \# of thread end & \\
         
         \bottomrule
    \end{tabular} 
    }
\end{table}

\subsubsection{Attack detection} \label{ssec:detection}


Peeler uses two classification models (MLR and SVM-RBF), and finally decides the classification outcome by fusing their scores. We note that MLR and SVM-RBF are constructed with different feature sets -- MLR is trained with $FV_{\text{MLR}}$ while SVM-RBF is trained $FV_{\text{SVM}}$ (see Table~\ref{tab:feature extraction}). The scores from the two models are fused by taking their average for detection.



\section{Dataset collection} \label{sec: dataset collection}

We aimed to collect a ransomware dataset containing diverse ransomware families rather than similar ransomware variants. Therefore, we collected ransomware samples from several sources including VirusTotal~\cite{virustotal}, malware repository~\cite{thezoomalware}, malwares~\cite{malware-samples}, and other online communities. Also, we collected benign applications exhibiting at least one of the following behaviors: 1) encryption or compression capabilities, 2) spawning multiple processes, and 3) most commonly used benign applications (see Appendix~\ref{appendix: benign applications}) to evaluate Peeler's robustness against false positives. 

\subsection{Ransomware} \label{ssec:ransomware dataset}

We collected 28,034 ransomware samples from VirusTotal~\cite{virustotal}, MalwareBazaar~\cite{malwarebazaar}, malware repository~\cite{thezoomalware}, malwares~\cite{malware-samples}, and other online communities. 
However, we excluded many malware samples from our final dataset for experiments. 
First, we found that many samples were not actual ransomware samples, although they were classified as ransomware by some vendors in VirusTotal. Therefore we discarded such samples. This finding is consistent with the observation in the previous work~\cite{scaife2016cryptolock}. 
Second, ransomware often needs to interact with command-and-control (C\&C) servers to perform their malicious activities. However, several ransomware samples did not often work appropriately because their corresponding C\&C servers were taken-down. Also, some sophisticated malware samples can detect the analysis environment and remain inactive to evade detection~\cite{miramirkhani2017spotless}. 
More importantly, samples from a few ransomware families we observed were significantly larger compared to other families. For example, we found more than 20,000 ransomware samples from Virlock family including \textit{Virlock Gen.1}, \textit{VirLock Gen.4}, and \textit{VirLock Gen.8} variants.
Therefore, we kept limited samples from Virlock family and discarded other samples. 
Finally, We collected 292 active samples from 67 ransomware families that perform their activities correctly.
We used 206 ransomware samples from 43 ransomware families (see Table~\ref{tab:ransomware families}) in the first set of experiments. Out of 43 families, 34 (102 samples) were from crypto, and 9 (104 samples) were from screen-locker types of ransomware. The remaining 24 ransomware families with 86 samples were collected at later stage of data collection and used to evaluate Peeler on new and unseen ransomware samples.

\begin{table}[h]
    \centering
    \caption{Ransomware families, types, and samples.} \label{tab:ransomware families}
    \resizebox{3.2in}{!}{
    \begin{tabular}{rllr|rllr}
    \toprule
         \textbf{no.} & \textbf{Family} & \textbf{Type} & \textbf{Samples} & \textbf{no.} & \textbf{Family} & \textbf{Type} & \textbf{Samples}\\
         \midrule
         1 & Cerber & Crypto & 33 & 23 & Petya & Crypto & 1 \\
         2 & Sodinokibi & Crypto & 14 & 24 & Satana & Crypto & 1 \\
         3 & GoldenEye & Crypto & 12 & 25 & Shade & Crypto & 1 \\
         4 & Sage & Crypto & 5 & 26 & Syrk & Crypto & 1 \\
         5 & Locky & Crypto & 5 & 27 & TeslaCrypt & Crypto & 1 \\
         6 & Dharma & Crypto & 3 & 28 & ucyLocker & Crypto & 1 \\
         7 & dotExe & Crypto & 3 & 29 & Unlock92 & Crypto & 1 \\
         8 & Troldesh & Crypto & 1 & 30 & Vipasana & Crypto & 1 \\
         9 & WannaCry & Crypto & 3 & 31 & Xorist & Crypto & 2 \\
         10 & Da Vinci Code & Crypto & 1 & 32 & Malevich & Crypto & 1 \\
         11 & CryptoShield & Crypto & 1 & 33 & Jigsaw & Crypto & 1 \\
         12 & CryptoWire & Crypto & 1 & 34 & Adobe & Crypto & 1\\
         13 & District & Crypto & 1 & 35 & Virlock.Gen.5 & Screen & 83 \\
         14 & Gandcrab & Crypto & 1 & 36 &  LockScreen.AGU & Screen & 12 \\
         15 & GlobeImposter & Crypto & 1 & 37 & Alphabet & Screen & 2 \\
         16 & Hexadecimal & Crypto & 1 & 38 & EgyptianGhosts & Screen & 1\\
         17 & InfinityCrypt & Crypto & 1 & 39 & Lockey-Pay & Screen & 1 \\
         18 & IS (Ordinpt) & Crypto & 1 & 40 & Blue-Howl & Screen & 1 \\
         19 & Keypass & Crypto & 1 & 41 & ShellLocker & Screen & 1 \\
         20 & Lockcrypt & Crypto & 1 & 42 & DerialLock & Screen & 1\\
         21 & Pack14 & Crypto & 1 & 43 & Trojan.Ransom & Screen & 1 \\ 
         22 & PocrimCrypt & Crypto & 1 & \multicolumn{4}{c}{-} \\ \midrule
         \multicolumn{8}{c}{\textbf{Total samples: 206, crypto = 102, screen-locker = 104}} \\
         \bottomrule
    \end{tabular} 
    }
\end{table}

\subsubsection{User environment and ground truth (labeled) dataset} 

We used VirtualBox 6.1~\cite{virtualbox} to create and manage the computing environment locally for experiments. Rather than using artificially generated data, we used a real user's data running on the Windows 10 64Bit operating system (a copied version of real user data) to set up a benign user's environment realistically. Multimedia files (e.g., \texttt{bmp}, \texttt{jpeg}, \texttt{png}, and \texttt{mp4}), Microsoft office documents (e.g., \texttt{docx}, \texttt{xlsx} and \texttt{pptx} files), and other important files (e.g., \texttt{cpp}, \texttt{py}, \texttt{pdf} and \texttt{wav} files) were copied to various directories in different locations. We note that those files are typically most attractive targets for ransomware. 


Each ransomware sample was executed and then manually labeled by each family type. We ran each ransomware sample for ten minutes or until all user files were encrypted. It took more than 90 days to run all samples and collect data. We only considered those samples that encrypted user files or locked desktop screens. If no files were modified, we excluded them from our dataset. 
We also obtained labeled ransomware samples from two well-known malware repositories~\cite{malwarebazaar,thezoomalware}. 
Many ransomware families use their respective names as file extensions after encrypting a user file, e.g., \textit{WannaCry} adds \textit{.wncry} file extension after encrypting a user file, similarly, \textit{Ranzy}, \textit{Ryuk}, and \textit{Peta} add \textit{.rnz}, \textit{.ryuk}, and \textit{.peta} file extensions, respectively. Moreover, we manually verified label ransomware family from VirusTotal's vendors, i.e., if more than 15 vendors assign the same label to a sample, we label it accordingly.

\subsubsection{Diversity in our dataset}

Table~\ref{tab:ransomware families} presents a list of ransomware families that are used in our evaluation. To the best of our knowledge, this is the most comprehensive dataset containing diverse ransomware families. According to previous work~\cite{nieuwenhuizen2017behavioural, scaife2016cryptolock}, the use of diverse families is more important than the number of ransomware samples from a few families for evaluating the performance of ransomware detectors. For instance, building a model on 1,000 Locky (and its variants) ransomware samples should prove no more useful than building a model on just one Locky sample~\cite{nieuwenhuizen2017behavioural}. 
Scaife et al.~\cite{scaife2016cryptolock} confirmed that due to the homogeneous nature of file I/O behavior among samples within each family, a small number of representative samples in each family are sufficient to evaluate the detection performance. It is because the core behavioral traits shown by crypto ransomware in encrypting data attack does not change from one variant to the other within a family. Since our study covered more than eight times the number of families from previous study~\cite{kharraz2015cutting}, and more than two times the number of families covered in studies~\cite{kharaz2016unveil, scaife2016cryptolock} and there was not much diversity within families, there was little need to collect additional samples.

\subsection{Benign applications} \label{ssec: benign user applications}

We also collected the dataset for popularly used applications that are typically installed on a benign user's computer. In addition to popularly used applications, we also considered several benign applications that could resemble ransomware in certain behavioral aspects. The reason is to investigate false positive rates when benign applications potentially resemble ransomware. We divide the benign dataset into three main categories targeting various types of ransomware: 1) benign applications with file I/O patterns resembling crypto ransomware, 2) benign applications spawning many processes, and 3) commonly used benign applications.

\subsubsection{Benign applications resembling crypto ransomware}

Benign applications performing encryption or compression might generate file I/O patterns similar to crypto ransomware that could result in false positives. To evaluate Peeler against, we collected data from several crypto-like benign applications, listed in Appendix~\ref{appendix: benign applications}. There are key differences in file I/O patterns generated by encryption/compression tools compared to crypto-ransomware. Firstly, unlike ransomware incurring a massive number of repeated file I/O patterns, the encryption/compression tools operate on a limited number of files only. Secondly, the original user files remain intact, i.e., not overwritten or deleted, even after compression or encryption is performed. It is, therefore, doubtful that benign applications show ransomware file I/O patterns. Thirdly, unlike crypto ransomware that encrypts user files arbitrarily, benign applications create a dedicated process that needs sophisticated inputs from the OS to complete the task. For instance, the compression tool \textit{7-zip} takes several parameters to specify target files. Each tool in Appendix~\ref{appendix: benign applications} is run twice -- for compression and decompression, on a given set of files to collect data.


\subsubsection{Benign applications spawning many processes}

We collected the dataset containing applications that spawn many child processes to evaluate Peeler's robustness against false positives. We found that certain benign applications such as Pycharm and Visual Studio create many spawned processes that may resemble screen-locker ransomware. The list of benign applications is shown in Appendix~\ref{appendix: benign applications}. We collected data by running each application individually. 


\subsubsection{Commonly used benign application}

We also collected user's system usage data under normal conditions while interacting with commonly used applications. A user runs many different applications at the same time. For example, the user read a document using Adobe Acrobat Reader, switched to the internet browser to view online reviews about a product, and then used Adobe Acrobat Reader again. Our goal here is to analyze system events generated in an interleaved manner from commonly used benign applications. The collected data is for around 12 hours of computer usage. During data collection, the user interacted with multiple applications from the list of benign applications in Appendix~\ref{appendix: benign applications}.


\section{Evaluation} \label{sec: evaluation}

We demonstrate Peeler's performance in detection accuracy, detection time, and CPU/memory overheads.  
For evaluation, we used the dataset described in Section~\ref{sec: dataset collection}. For training, 20\% of both screen-locker ransomware and benign applications randomly are selected. We used a small training dataset (only 20\% of the entire dataset) for the following reasons: 1) the size of a training set is typically limited in the real world; 2) we want to increase the number of testing samples as many as possible for making Peeler robust against unseen ransomware samples; 3) we want to reduce the size of model for quick training. All the remaining ransomware and benign samples (i.e., 80\% of the total samples) are used for testing purposes.

\subsection{Detection accuracy} \label{ssec: ransomware detection results}

Table~\ref{tab:evaluation results} shows the summary of Peeler's detection accuracy. Overall, Peeler achieved 99.52\% accuracy with a false positive rate of only 0.58\%. Since we used only 20\% of applications selected randomly for training, all performance statistics are averaged after 100 runs to mitigate biases in training and testing datasets. Similarly, Peeler achieved high precision and F1 score, which are greater than 99\%. 

\begin{table}[h]
    \centering
    \caption{Peeler's detection accuracy.} \label{tab:evaluation results} 
    \resizebox{3.0in}{!}{
    \begin{tabular}{rrrrrrr}
    \toprule
         \textbf{Acc. (\%)} & \textbf{TPR (\%)} &\textbf{FPR (\%)} &\textbf{FNR (\%)} & \textbf{Prec. (\%)} & \textbf{Rec. (\%)} & \textbf{F1 (\%)}\\ \midrule
         99.52 & 99.63 & 0.58 & 0.37 & 99.41 & 99.63 & 99.52 \\
         \bottomrule
    \end{tabular}
    }
\end{table}



\subsubsection{False positive analysis} \label{ssec:false positives analysis}

Minimizing false positives is essential to develop practically useful malware detectors because excessive false positives can annoy users and undermine the system's effectiveness. We evaluate Peeler's performance against three different types of benign application scenarios (see Table~\ref{tab: benign applications false positive rates}):

\begin{table}[h]
    \centering
    \caption{Peeler's false positive analysis.} \label{tab: benign applications false positive rates}
    \resizebox{2.6in}{!}{
    \begin{tabular}{lrrr}
    \toprule
         \textbf{Scenario} & \textbf{TNR (\%)} &\textbf{FPR (\%)} &\textbf{FNR (\%)} \\ \midrule
         Crypto-like benign apps & 98.27& 1.72 & 0.96\\
         Locker-like benign apps & 99.5& 0.31 & 0.5\\
        Commonly used benign apps & 99.78& 0.21 & 0.87 \\ \midrule
         All ransomware & 99.42& 0.58 & 0.37\\
         \bottomrule
    \end{tabular}
    }
\end{table}


\textbf{Crypto ransomware-like benign applications.} For behavior-based crypto ransomware detection solutions, a significant challenge is not to detect benign applications having compression or encryption capabilities because their system behaviors might be similar to crypto ransomware. 

We deeply investigated 11 different applications using compression or encryption operations on a large number of files like crypto ransomware (see Appendix~\ref{appendix: benign applications}). We observed that event sequences of some benign applications such as ZipExtractor and BreeZip are quite similar to those of typical crypto ransomware, but they do not restrict access to files via encryption, unlike crypto ransomware. 



Table~\ref{tab: benign applications false positive rates} shows that Peeler correctly detects 98.27\% with a false positive rate of 1.72\% even against crypto ransomware-like benign applications.

\textbf{Benign applications spawning many processes.} Certain ransomware spawns many child processes. Therefore, we examine how Peeler's performance can be degraded with benign apps having such behaviors. For this analysis, we investigated 34 most popular applications from Microsoft's Windows OS Store (\url{https://www.microsoft.com/en-us/store/apps/windows}) (see Appendix~\ref{appendix: benign applications}) and selected 18 applications showing such behaviors. Table~\ref{tab: benign applications with many spawned processes} presents the applications' process tree-related feature ($FV_{MLR}$) values of three representative benign applications showing such behaviors. We observe that Pycharm and Visual Studio spawned 140 and 46 child processes, respectively. Interestingly, Chrome spawns 42 processes, but the depth of its applications' tree is one, and the number of threads created by these processes is 1,480. We examined the results of Peeler with those 18 applications.  


\begin{table}[h]
    \centering
    \caption{Benign applications' process tree features.} \label{tab: benign applications with many spawned processes}
    \resizebox{3.4in}{!}{
    \begin{tabular}{lrrrrr}
    \toprule
         \textbf{Application} & \textbf{\# processes} & \textbf{Depth} & \textbf{\# leaf nodes} &\textbf{\# unique processes} & \textbf{\# threads}\\ \midrule
         Pycharm & 140 & 4 & 70 & 11 & 993 \\
         Visual Studio & 46 & 4 & 29 &21&568 \\
         Chrome & 42 & 1& 41& 2& 1,480 \\
         \bottomrule
    \end{tabular}
    }
\end{table}

Table~\ref{tab: benign applications false positive rates} shows that Peeler correctly detected 99.5\% with a false positive rate of 0.31\% even though these benign applications spawned many processes. The reason for this detection is because Peeler also considers the other set of features ($FV_{SVM}$), which are related to system events.

\textbf{Commonly used benign applications.} We also evaluated Peeler's performance with commonly used benign applications such as Microsoft Office, Adobe Acrobat Reader, email client, and instant messengers, as presented in Sectoin~\ref{ssec: benign user applications}. 




We show that Peeler correctly detects all benign activities performed by a user achieving a detection rate of 99.78\%. The false positive and true negative rates under normal system usage are 0.21\% and 0.87\%, respectively. Note that the overall detection accuracy in all three scenarios is above 99\%.






\subsubsection{False negative analysis} \label{ssec:false negative analysis}

The false negative rate for ransomware detection is another important metric to evaluate the effectiveness of Peeler. Table~\ref{tab: malicious applications false negative rates} shows that the overall false negative rate is above 0.37\%. The false negative rates for crypto and screen-locker ransomware are 0\% and 0.5\%, respectively.



\begin{table}[h]
    \centering
    \caption{Peeler's false negative rate analysis.} \label{tab: malicious applications false negative rates}
    \resizebox{2.0in}{!}{
    \begin{tabular}{lrrr}
    \toprule
         \textbf{Ransomware} & \textbf{TPR (\%)} &\textbf{FPR (\%)} &\textbf{FNR (\%)} \\ \midrule
         Crypto & 100& 0.8 &0\\ 
         Screen-locker &99.50 & 0.31 & 0.50 \\\midrule
         All ransomware &99.63&0.58&0.37 \\
         \bottomrule
    \end{tabular}
    }
\end{table}


\subsection{Detection time} \label{ssec:early detection}
For crypto ransomware, the detection time refers to the time interval between the end of the detection and the end of encryption on a file by a process, that is, how long it takes Peeler to detect the ransomware attack after a ransom sample encrypts a file. We excluded the time to be taken for the file encryption because that time can be varied depending on the file size.

\begin{figure}[!ht]
  \centering
  \includegraphics[width=2.7in,clip]{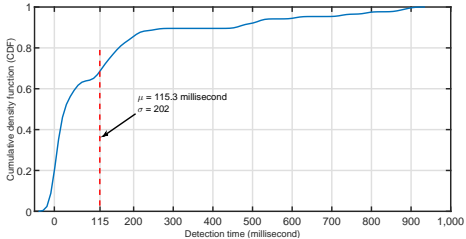}
  \caption{CDF of the crypto ransomware detection time.}
  \label{fig:timeresultscrypto}
\end{figure}

Figure~\ref{fig:timeresultscrypto} shows that Peeler can detect over 70\% of samples within 115 milliseconds with the mean time of 115.3 milliseconds, demonstrating that Peeler outperforms existing crypto ransomware detection solutions in detection time. Peeler can promptly detect crypto ransomware with a simple, file I/O pattern matching, unlike other existing solutions relying on complicated file activity (e.g., identifying encryption operations based on entropy computation) analysis or machine learning models. 

These detection time results demonstrate the superiority of Peeler compared with existing crypto ransomware detection solutions. Cryptolock~\cite{scaife2016cryptolock} detects crypto ransomware after 10 files are encrypted. Similarly, Mehnaz et al.~\cite{mehnaz2018rwguard} presented a solution which takes on average 8.87 seconds to detect malicious processes.

\begin{figure}[!ht]
  \centering
  \includegraphics[width=2.8in,clip]{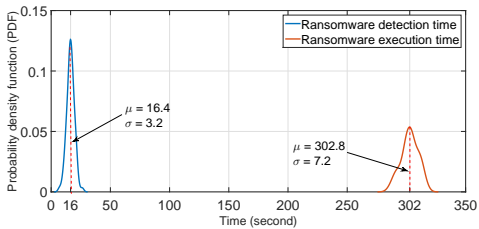}
  \caption{Probability density function of detection times for screen-locker ransomware with mean ($\mu$) and standard deviation ($\sigma$).}\label{fig:timeresults}
\end{figure}

For screen-locker ransomware, the detection time refers to the time interval between the end of the detection and the start activity by a ransomware process. Peeler, on average, took 16.4 seconds to detect screen-locker ransomware while the execution time of screen-locker ransomware to lock user's desktop screen completely is 302.8 seconds on average, demonstrating that Peeler can detect screen-locker ransomware at a very early stage that prevents an attacker from locking a victim's system. Figure~\ref{fig:timeresults} shows the probability density function of screen-locker detection time. Moreover, the detection time is not affected even if ransomware is running in parallel to many other irrelevant system applications because Peeler's system events monitor directly intercept the events and then analyse them.

\subsection{Robustness against unseen families} \label{ssec: new samples detection}

To test Peeler against unseen ransomware families, we additionally collected new and \textit{unseen} ransomware samples after three months from the first experiments and monitored online repositories for new or unseen ransomware samples. A total of 70 samples from more than 24 distinct unseen or new ransomware families are tested. We used the previously constructed Peeler without retraining. All samples tested in this experiment are manually verified from VirusTotal and other online malware repositories to confirm their family and type.

Peeler was able to correctly detect 67 samples from total 70 new and unseen ransomware samples achieving more than 95\% detection rate. The ransomware families and the number of corresponding samples tested are given in Table~\ref{tab:unseen families}.

Peeler can also detect 9 out of 16 conventional malware samples even though those samples do not have ransomware capabilities. For example, our investigation revealed that \textit{Backdoor.Generic} is a malware family that enables attackers to control infected computers remotely to create large groups of zombie computers (botnets), which are then used for malicious purposes without the user's knowledge. We surmise that those malware samples have some common behaviors that can be observed from ransomware.

\begin{table}[h]
\centering
\caption{Detection of \textit{unseen} ransomware and malware.} \label{tab:unseen families}
\resizebox{2.8in}{!}{
\begin{tabular}{llrlr}
\toprule
\textbf{\makecell{Type/\\Detection}} & \textbf{Family} & \textbf{Sample} & \textbf{Family} & \textbf{Sample}\\ \toprule
\multirow{12}{*}{\makecell{Ransomware \\ 95.7\% (67/70)}} & Ryuk &  6 & Zeppelin & 6  \\ 
& Ranzy & 4 & Netwalker & 2  \\
&Core & 3 &  Fox & 3   \\
&Crpren & 1 & MedusaLocker & 1 \\ 
&Balaclava & 5 & Crylock & 7 \\
&Matrix & 4 & DarkSide & 4 \\
&Ragnarlocker & 2 & HiddenTear & 2 \\
&Mespinoza & 5 & Thanos & 3  \\
&Vaggen & 3 & Mountlocker & 2 \\
&Nemty & 2 & Phobos & 1  \\
&Jsworm & 1 & Winlock & 1  \\
&Maze & 1 & Unknown & 1 \\ \midrule
{\makecell{Malware \\ 56.2\% (9/16)}} & Backdoor.Generic& 2 &  Unknown & 14 \\ \bottomrule
\end{tabular}}
\end{table}

\subsection{Model-specific detection accuracy.}
We constructed Peeler by building multiple detection components (file I/O pattern matcher and machine learning-based classifier). Here we show the effectiveness of each component in detecting ransomware. Table~\ref{tab:component1 results} and~\ref{tab:component2 results} show the evaluation results of both components, respectively. 

\begin{table}[h]
    \centering
    \caption{File I/O pattern matcher's performance.} \label{tab:component1 results}
    \resizebox{3.2in}{!}{
    \begin{tabular}{lrrrrrr}
    \toprule
         \textbf{Ransomware} & \textbf{TPR (\%)} &\textbf{FPR (\%)} &\textbf{FNR (\%)} & \textbf{Prec. (\%)} & \textbf{Rec. (\%)} &\textbf{F1 score (\%)} \\ \midrule 
         Crypto & 95.19 & 0 & 4.81 & 100 & 95.19 & 97.53\\ 
         Screen-locker &33.33 & 0 & 66.66 & 100 & 33.33 & 50.00 \\ \midrule
         All ransomware & 65.50 & 0 & 34.5 & 100 & 65.50 & 79.15 \\
         \bottomrule
    \end{tabular}
    }
\end{table}

Table~\ref{tab:component1 results} shows that the file I/O pattern matcher achieves the detection rate of 95.19\% with the false rate of 0\% in detecting crypto ransomware; however, it is not effective in detecting screen-locker ransomware. 

Table~\ref{tab:component2 results} shows that the machine learning-based classifier works well in detecting both crypto and screen-locker ransomware. The machine learning-based classifier achieves more than 98\% detection accuracy with less than 2\% false rate in detecting both types of ransomware attacks. The high detection rate of the machine learning-based classifier can be attributed to the feature vectors ($FV_{\text{MLR}}$ and $FV_{\text{SVM}}$). However, to boost the detection time, we can first apply the file I/O pattern matcher and then use the machine learning-based classifier only when the file I/O pattern matcher fails to detect.

\begin{table}[h]
    \centering
    \caption{Machine learning-based classifier's performance.} \label{tab:component2 results}
    \resizebox{3.2in}{!}{
    \begin{tabular}{lrrrrrr}
    \toprule
         \textbf{Ransomware} & \textbf{TPR (\%)} &\textbf{FPR (\%)} &\textbf{FNR (\%)} & \textbf{Prec. (\%)} & \textbf{Rec. (\%)} &\textbf{F1 score (\%)} \\ \midrule 
         Crypto & 98.45 & 1.84 & 1.54 & 98.24 & 98.45 & 98.29\\ 
         Screen-locker &99.5 & 0.31 & 0.50 & 99.68 & 99.50 & 99.59 \\ \midrule
         All ransomware & 99.05 & 1.9 & 0.94 & 98.19 & 99.05 & 98.59 \\
         \bottomrule
    \end{tabular}
    }
\end{table}

Each component can work as an alternative and complementary detection method to the other component. For instance, the file I/O pattern matcher failed to detect some crypto ransomware samples such as Hexadecimal, Cryptowire, CryptoShield, and CryptoLock, but the machine learning-based classifier successfully detected them. 





\subsection{CPU and memory overheads} \label{ssec:performance overhead}
We evaluate the performance of Peeler with respect to CPU and memory overheads. Since Peeler intercepts low-level kernel events and then analyzes them to detect ransomware attacks, its performance overheads in CPU and memory can typically be changed depending on the system's workload. For instance, we observe that if computationally intensive tasks are running, the overheads of Peeler inherently increase because Peeler should frequently intercept a high number of kernel events generated from such computationally intensive processes.
\begin{table}[h]
\centering
\caption{CPU and memory overheads of Peeler.} \label{tab:overhead}
\resizebox{1.7in}{!}{
\begin{tabular}{l|rr|rr}
\toprule
\multirow{2}{*}{\textbf{Workload}} & \multicolumn{2}{c|}{\textbf{CPU (\%)}} & \multicolumn{2}{c}{\textbf{Memory (MB)}} \\  \cmidrule{2-5}
& Mean& Std.& Mean & Std.\\ \midrule
Low & 2.7 & 2.0 & 9.8 & 0.3 \\ \midrule
Normal & 4.9 & 2.2 & 9.8 & 0.4 \\ \midrule
High & 15.8 & 3.9 & 11.3 & 1.9 \\ \bottomrule
\end{tabular}}
\end{table}

Our experiments were conducted on a computing device equipped with two Intel Core(TM) i5-7300U (2.60GHz) CPUs and 8GB RAM, running 64-bit Windows 10 Enterprise edition operating system. 
Our CPU and memory usage results were measured based on this environment. We considered the three workload conditions as follows: Low, Normal, and High workload conditions. 
Under the low workload condition, Peeler is only running in the background and continuously collecting system events and writing them to a log file. Under the normal workload condition, we additionally performed the following tasks: 
1) drafting an MS Word document; 
2) using Chrome for browsing online material; and 
3) reading a document using Adobe Acrobat Reader. Under the high workload condition, we additionally ran a CPU intensive algorithm as a background process.
We note that an anti-malware program called CylanceProtect and default Windows OS services were continuously running in the background for all workload conditions.

Table~\ref{tab:overhead} shows CPU and memory usage of Peeler. We observe that mean CPU usage and the standard deviation are 4.9\% and 2.2\%, respectively, under the normal workload condition.

Table~\ref{tab:overhead} also shows that mean memory usage is around 9.8MB under the low or normal workload conditions, which is quite stable and has less variation (standard deviation is 0.4MB). Under the high workload condition, the average CPU usage of Peeler significantly increases to 15.8\% with a standard deviation of 3.9\% while its average memory usage slightly increases to 11.3MB memory with a standard deviation of 1.9MB. 


\section{Discussion and limitations} \label{sec:discussion and limitations}
\subsection{Comparison with existing solutions}
As mentioned in Section~\ref{sec:introduction}, there are many existing methods to detect ransomware attacks. However, since most existing solutions used their own dataset for evaluation, and their source code is not opened, we do not directly compare Peeler with those solutions. Alternatively, we compare Peeler with those solutions according to their experimental results reported in their papers. Table~\ref{tab:comparison} shows a summary of the comparison results.

\begin{table}[!ht]
\centering
\caption{Comparison with existing approaches.} \label{tab:comparison}
\resizebox{3.5in}{!}{
\begin{tabular}{lrrrcrc}
\toprule
\textbf{Method} & \textbf{TPR (\%)} & \textbf{FPR (\%)} & \textbf{Files lost} & \textbf{Screen-locker?} &\textbf{Samples/families} &\textbf{Real-time} \\ \midrule
Redemption~\cite{kharraz2017redemption} & 100 & 0.8 & 5 & $\times$ & 677/29& $\times$ \\
CryptoLock~\cite{scaife2016cryptolock} & 100 & 0.03 & 10 & $\times$ & 492/14 & $\times$\\
UNVEIL~\cite{kharaz2016unveil} & 96.3 & 0 & - & $\checkmark$ &  2121/ -&  $\times$ \\
REDFISH~\cite{morato2018ransomware}& 100 & - & 10 & $\times$ & 54/19& $\times$ \\
RWGuard~\cite{mehnaz2018rwguard} & - & 0.1 & partial recovery & $\times$ & - /14& $\checkmark$ \\
Elderan~\cite{sgandurra2016automated} & 93.3 & 1.6 & - & $\times$ & 582/11& $\times$ \\
CM\&CB~\cite{ahmadian2015connection}& 98 & Vary & - & $\times$ & 8/ -& $\times$ \\
RansHunt~\cite{hasan2017ranshunt}& 97 & 3 & - & $\times$ & 360/20& $\times$ \\
ShieldFS~\cite{continella2016shieldfs} & 100 & 0.038 & - & $\times$ & 383/11& $\times$ \\
\textbf{Peeler} & \textbf{99.63} & \textbf{0.58} & \textbf{1} & $\checkmark$ & \textbf{206/43}& $\checkmark$ \\
\bottomrule
\end{tabular}}
\end{table}

We can see that the ransomware detection rates overall ranges from 93\% to 100\%. However, it is important to note that the number of ransomware samples/families used for evaluation is different for each approach. CryptoLock~\cite{scaife2016cryptolock}, Redemption~\cite{kharraz2017redemption}, REDFISH~\cite{morato2018ransomware}, and ShieldFS~\cite{continella2016shieldfs} achieved 100\% detection rates, but those solutions were tested on 14, 29, 19, and 11 ransomware families only, respectively. In contrast, Peeler was tested against 43 distinct ransomware families, including both crypto and screen locker ransomware, and still achieved a 99.63\% detection rate. For the false positive rate with benign applications, Peeler achieved 0.58\% FPR, which would be comparable with the other solutions.
For ransomware detection, one of the most critical evaluation metrics is the number of user files lost before a ransomware sample is detected. Peeler detects ransomware immediately after a single file alone is encrypted, indicating that Peeler outperforms other solutions.
From Table~\ref{tab:comparison}, we can see that only two solutions (Peeler and UNVEIL~\cite{kharaz2016unveil}) considered the detection of screen-locker ransomware. However, Peeler's detection time (16.4 seconds on average) would significantly be faster than UNVEIL. Although Kharaz et al.~\cite{kharaz2016unveil} did not report the exact detection time of UNVEIL, we surmise that UNVEIL would take longer detection time because it needs to capture screenshots of a victim's desktop periodically, and then compute the similarity between the captured images.

\subsection{Secure implementation of Peeler}

Peeler runs in a privileged kernel mode with administrative rights because it needs to collect kernel-level events from four main kernel providers (\texttt{File}, \texttt{Process}, \texttt{Image}, and \texttt{Thread}). Therefore, the integrity of Peeler can be securely protected against malicious processes running in the user mode. However, if we consider powerful attackers (e.g., rootkit-based ransomware) with the root privilege, we additionally need to consider deploying a kernel protection mechanism to protect Peeler against attackers. Recently, several techniques (e.g., real-time kernel protection (RKP)~\cite{azab2014hypervision}) have been proposed to protect the kernel code and objects from unauthorized modifications. With such a secure kernel protection mechanism, we can implement Peeler as a Windows OS service running on the kernel.

\subsection{Peeler's extension to other computing environments}


In order to extend Peeler to other computing environments such as Linux or Android, two aspects are needed to be considered: platform-depended system events and key ransomware behavioral characteristics. The later remains platform-independent because the key behavioral characteristics of ransomware remain almost the same no matter which platform they target. However, system events for other computing environments need to be carefully analyzed to extend Peeler to these environments. For instance, the suspicious file I/O patterns and corresponding regular expressions should be updated to reflect platform-dependent system events.

\subsection{Limitations}

In our experiments, the tested benign applications are only a part of a massive number of benign applications. Therefore, we can still have a chance to encounter (unknown) benign applications that generate suspicious file I/O patterns that might result in false positives. For instance, when Windows OS creates a backup file \texttt{tokens.dat.bak} from the file of \texttt{tokens.dat}, we found that the sequence of kernel-level file I/O events generated from a benign process can lead to a false positive result from the file I/O pattern matcher. 

A sophisticated malware could memory map a user file and then copy the block of memory, using \textit{memcpy()} to evade file I/O patterns. The Windows OS may not observe any write operation in such scenario, and therefore, Peeler may not detect such attacks.

\section{Related work} \label{sec:related work}
We categorize the literature regarding ransomware detection into three groups: 1) crypto ransomware detection techniques that are mainly based on certain behavioral indicators (e.g., file I/O event patterns), 2) machine learning-based approaches that build models by leveraging system behavior feature, and 3) decoy-based approaches that deploy decoy files and monitor if ransomware samples to tamper with the decoy files.

\textbf{Crypto ransomware detection.} There were several proposals to monitor file I/O request patterns of applications to detect crypto ransomware. Kharraz et al.~\cite{kharaz2016unveil} studied crypto ransomware families' file I/O request patterns and presented a dynamic analysis-based ransomware detection system called UNVEIL. UNVEIL detected 13,647 ransomware samples from a dataset of 148,223 general malware samples. Kharraz et al.~\cite{kharraz2017redemption} proposed another ransomware detection system using file I/O patterns, achieving a 100\% detection rate with 0.8\% false positive on 677 samples from 29 ransomware families. Scaife et al.~\cite{scaife2016cryptolock} also presented a system called CryptoDrop that detect ransomware based on suspicious file activities, e.g., tampering with a large number of file accesses within a time interval. According to the experimental results, the number of lost files is ten on average. Moratto et al.~\cite{morato2018ransomware} proposed a ransomware detection algorithm with a copy of the network traffic, without impacting a user's activities. Their proposed system achieved a 100\% detection rate with 19 different ransomware families after the loss of ten files. 

\textbf{Machine learning based ransomware detection.} RWGuard~\cite{mehnaz2018rwguard} is a machine learning-based crypto ransomware detection system. It achieved a 0.1\% false positive rate incurring a 1.9\% CPU overhead with 14 crypto ransomware families. RWGuard leverages the features about processes' I/O requests and changes performed on files because RWGuard was mainly designed to detect crypto ransomware only. 
Sgandurra et al.~\cite{sgandurra2016automated} proposed EldeRan, another machine learning approach that builds a model using system activities such as API invocations, registry event, and file operations that are performed by applications. EldeRan achieved a 93.3\% detection rate with a 1.6\% false alarm rate with 582 samples from 11 ransomware families. Hirano et al.~\cite{hirano2019machine} and Al-rimy~\cite{al20170} proposed behavior-based machine learning models for ransomware detection. Hirano et al. selected five-dimensional features that were extracted both from ransomware and benign applications' I/O log files. Nieuwenhuizen~\cite{nieuwenhuizen2017behavioural} proposed another behavioral-based machine learning model using a feature set that quantifies the behavioral traits for ransomware's malicious activities. 

\textbf{Decoy files based ransomware detection.} Decoy techniques~\cite{mehnaz2018rwguard,gomez2018r,continella2016shieldfs} have also been frequently proposed to detect ransomware attacks. For example, Gomez et al.~\cite{gomez2018r} developed a tool called R-Locker using honey files to trap the ransomware. When file operations are performed on honey files by a process, the process is detected and completely blocked because benign processes do not perform any file operations on honey files. However, if decoy files are generated, which look different from real user files, sophisticated ransomware samples ignore decoy files~\cite{mehnaz2018rwguard}. Moreover, it is also unclear how those solutions would detect some ransomware families (e.g., Petya) that affect predefined system files only.

\section{Conclusion} \label{sec:conclusion and future work}
We propose a new effective and efficient solution called Peeler to detect ransomware attacks using their system behaviors. Peeler is built on both rule-based detection (e.g., malicious commands detector and I/O pattern matcher) and machine learning models to improve the detection accuracy and reduce the detection time. Most crypto ransomware can be detected by the I/O pattern matcher efficiently; the crypto ransomware that cannot be detected by the I/O pattern matcher or screen-locker ransomware can be detected by the machine learning models more accurately.

To show the effectiveness of Peeler, we evaluate its performance with 43 ransomware families containing crypto ransomware and screen-locker ransomware. In the experiments, Peeler achieved 99.52\% accuracy with a false positive rate of only 0.58\%. Moreover, Peeler is efficient in detecting crypto ransomware -- over 70\% of crypto ransomware samples can be detected within 115 milliseconds. Although Peeler's detection time (16.4 seconds on average) is relatively slower for screen-locker ransomware, it is still sufficient to detect it because it typically takes a longer time (302.8 seconds on average) to lock a victim's system entirely.



\bibliographystyle{ieeetr}
\bibliography{bibliography}
\newpage
\appendix
\subsection{File encryption patterns} \label{appendix: file encryption patterns}

\begin{figure}[!ht]
  \centering
  \includegraphics[width=3.4in,clip]{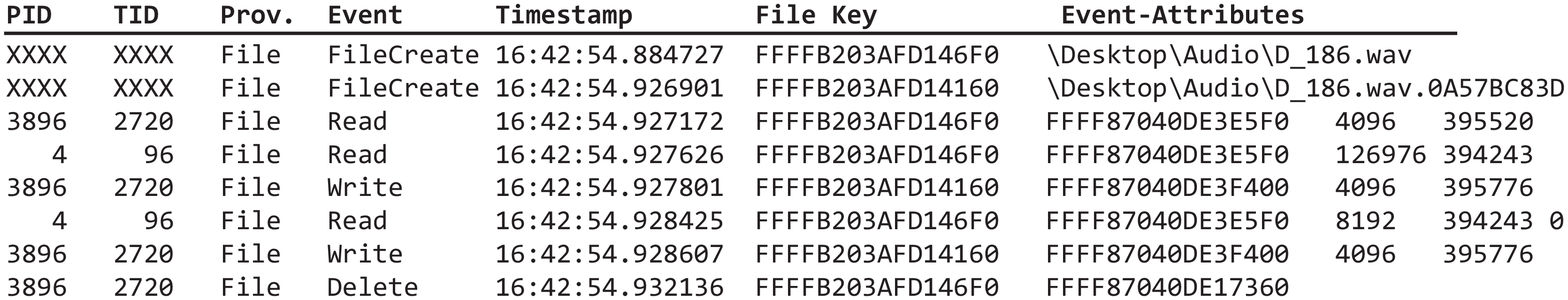}
  \caption{File I/O events generated by InfinityCrypt ransomware.}\label{fig:strat2}
\end{figure}

\begin{figure}[!ht]
  \centering
  \includegraphics[width=3.4in,clip]{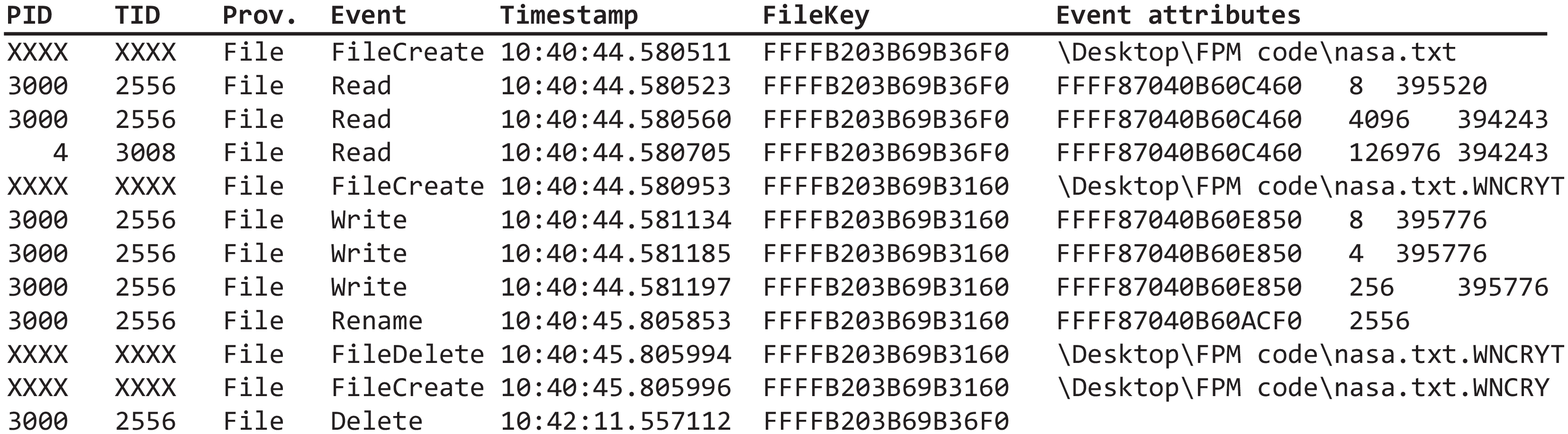}
  \caption{File I/O events generated by Wannacry ransomware.}\label{fig:strat3}
\end{figure}

\subsection{Benign applications} \label{appendix: benign applications}
In this section, we present benign applications that potentially show ransomware-like behavior that are used in evaluation of Peeler: 1) benign encryption, compression and shredder applications (Table~\ref{tab: resembles crypto ransomware}), 2) benign applications most commonly used, and 3) benign application spawning multiple processes (Table~\ref{tab: resembles locker ransomware}).
\begin{table}[h]
    \centering
    \caption{Ransomware-like benign applications.} \label{tab: resembles crypto ransomware}
    \resizebox{2.4in}{!}{
    \begin{tabular}{l|llr}
    \toprule
         \textbf{Tool} & \textbf{Application} & \textbf{Operation} & \textbf{Version}\\
         \midrule
         \multirow{12}{*}{Compression} & 7-zip & Compression & \multirow{2}{*}{19.00} \\
         &7-zip & Decompression & \\ 
         &Winzip & Compression & \multirow{2}{*}{24} \\
         &Winzip & Decompression & \\ 
         &Winrar & Compression &  \multirow{2}{*}{5.80} \\
         &Winrar & Decompression & \\ 
         &BreeZip & Compression &  \multirow{2}{*}{-} \\
         &BreeZip & Decompression & \\ 
         &Alzip & Compression &  \multirow{2}{*}{11.04}\\
         &Alzip & Decompression &\\ 
         &PeazipWinrar & Compression &  \multirow{2}{*}{7.1.1}\\
         &PeazipWinrar & Decompression &  \\ \midrule
         \multirow{4}{*}{Encryption} & AESCrypt & Encryption &  \multirow{2}{*}{310.00} \\
         &AESCrypt & Decryption & \\ 
         &AxCrypt & Encryption &  \multirow{2}{*}{-}\\
         &AxCrypt & Decryption &  \\ \midrule
         \multirow{3}{*}{Shredder} & Eraser & Delete &  6.2.0.2986\\
         &Ccleaner & Delete & - \\
         &Windows Delete & Delete & - \\
         \bottomrule
    \end{tabular}
    }
\end{table}

\begin{table}[h]
    \centering
    \caption{Benign applications spawning multiple processes.} \label{tab: resembles locker ransomware}
    \resizebox{3.2in}{!}{
    \begin{tabular}{clrc}
    \toprule
         \textbf{Type} & \textbf{Application} & \textbf{Version} & \textbf{spawn processes?}\\
         \midrule
         \multirow{5}{*}{Office} & MS Word & 16.0.11929.20436& $\times$ \\
         & MS Powerpoint & 16.0.11929.20436 &  $\times$\\
         & MS Excel & 16.0.11929.20436 & $\times$\\
         & MS Outlook & 16.0.11929.20436 & \checkmark\\
         & Trio Office: Word, Slide, Spreadsheet & - & \checkmark\\
         \midrule 
         \multirow{4}{*}{Development} & Pycharm & 11.0.3+12-b304.56 amd64 & \checkmark\\
         & Matlab & R2019a & \checkmark\\
         & Visual Studio C++ & 2019 community version & \checkmark\\
         & Android Studio & 191.6010548 & \checkmark \\
         \midrule
         \multirow{6}{*}{Tools} & Adobe Acrobat Reader & 20.006.20034 & \checkmark \\
         & Adobe Photoshop Express & 3.0.316 &$\times$\\
         & PhotoScape & 3.7 &$\times$\\
         & Cool File Viewer & - & $\times$\\
         & PicArt Photo Studio & - &$\times$\\
         & Paint 3D & - &$\times$\\
         \midrule
         \multirow{5}{*}{Cloud and Internet} & Dropbox & - &$\times$\\
         & Googledrive & - &$\times$\\
          & Internet Explorer & 11.1039.17763 & \checkmark\\
          & Google Chrome & 80.0.3987.132 & \checkmark\\
          & Remote Desktop & - & $\times$\\
          \midrule
          \multirow{4}{*}{Messenger} & Telegram & 1.9.7 &$\times$\\
          & WhatApp & 0.4.930 & \checkmark\\
          & Skype & 1.9.7 & $\times$\\
          & Facebook Messenger &  - &\checkmark\\
          \midrule
          \multirow{3}{*}{Document} & Wordpad & - &$\times$ \\
          & Notepad & - & $\times$\\
          & OneNote & 16001.12527.20128.0 &$\times$ \\
          \midrule
          \multirow{3 }{*}{Media player} & VLC & 3.0.8 & $\times$\\
          & Netflix & 6.95.602 & $\times$\\
          & GOM Player & 2.3.49.5312 & $\times$\\ 
          \midrule
          \multirow{4}{*}{Miscellaneous}  & Spotify & - & \checkmark \\
          & KeePass Password manager & 1.38 & $\times$\\
          & Discord & - & $\times$\\
          & Facebook & - & $\times$\\
         \bottomrule
    \end{tabular}
    }
\end{table}

\subsection{Malicious commands triggered during ransomware execution} \label{appendix: malicious commands}
In Table~\ref{tab:maliciouscommands}, we show example malicious commands extracted by Peeler during ransomware execution. This is not an exhaustive list of malicious commands instead we show few important malicious commands found during ransomware execution.
\begin{table}[t]
    \centering
    \caption{Malicious commands.} \label{tab:maliciouscommands}
    \resizebox{3.4in}{!}{
    \begin{tabular}{|r| p{8.4cm}|}
    \toprule
         \textbf{no} & \textbf{Command}\\
         \midrule
         \texttt{1} & \texttt{vssadmin.exe delete shadows /all /quiet}\\ \hline
         \texttt{2} & \texttt{bcdedit.exe /set {default} recoveryenabled No}\\ \hline
         \texttt{3} & \texttt{bcdedit.exe /set {default} bootstatuspolicy ignoreallfailures}\\ \hline
         \texttt{4} & \texttt{powershell.exe -e Get \-WmiObject Win32\_Shadowcopy | ForEach-Object\{\$\_.Delete();\}}\\ \hline
         \texttt{5} & \texttt{taskkill /t /f /im mal.exe} \\ \hline
         \texttt{6} & \texttt{del mal.exe} \\ \hline
         \texttt{7} & \texttt{reg add HKCU $\backslash$Software$\backslash$Microsoft$\backslash$Windows$\backslash$\newline  CurrentVersion$\backslash$Explorer$\backslash$Advanced /f /v HideFileExt /t REG\_DWORD /d 1}\\ \hline
        \texttt{8} &  \texttt{reg add HKCU $\backslash$Software$\backslash$Microsoft$\backslash$Windows$\backslash$\newline CurrentVersion$\backslash$Explorer$\backslash$Advanced /f /v Hidden /t REG\_DWORD /d 2}\\ \hline
        \texttt{9} & \texttt{reg add HKCU $\backslash$Software$\backslash$Microsoft$\backslash$Windows$\backslash$\newline CurrentVersion$\backslash$Explorer$\backslash$Advanced /f /v EnableLUA /d 0 /tREG\_DWORD /f} \\ \hline
        \texttt{10} & \texttt{reg add HKCU$\backslash$Control Panel$\backslash$Desktop /v Wallpaper /t REG\_SZ /d PathtoRansomNoteImage /f } \\ \hline
        \texttt{11} & \texttt{reg add HKCU$\backslash$Control Panel$\backslash$Desktop /v WallpaperStyle /t REG\_SZ /d "0" /f} \\ \hline
        \texttt{12} & \texttt{reg add HKCU$\backslash$Control Panel$\backslash$Desktop /v TileWallpaper /t REG\_SZ /d "0" /f} \\ \hline
        \texttt{13} & \texttt{powershell.exe -ExecutionPolicy Restricted -Command Write-Host 'Final result: 1';} \\ \hline
        \texttt{14} & \texttt{powershell.exe Set-MpPreference -DisableArchiveScanning \$true;} \\ \hline
        \texttt{15} & \texttt{powershell.exe Set-MpPreference -DisableBlockAtFirstSeen \$true;} \\ \hline
        \texttt{16} & \texttt{icacls Path /deny *S-1-1-0:(OI)(CI)(DE,DC)} \\ \hline
        \texttt{17} & \texttt{icacls . /grant Everyone:F /T /C /Q} \\ \hline
        \texttt{18} & \texttt{notepad.exe C:$\backslash$Users$\backslash$USER$\backslash$Music$\backslash$\# RESTORING FILES \#.TXT} \\ \hline
        \texttt{19} & \texttt{cscript  C:$\backslash$Users$\backslash$kim105$\backslash$AppData$\backslash$Local$\backslash$Temp/SUwk.vbs} \\ \hline
        \texttt{20} & \texttt{wmic.exe shadowcopy delete} \\ \hline
        \texttt{21} & \texttt{net.exe stop vss} \\ \hline
        \texttt{22} & \texttt{net.exe stop McAfeeDLPAgentService /y}\\ \hline
        \texttt{23} & \texttt{schtasks  /create /sc onlogon /tn TASK\_NAME /rl highest /tr PATH\_TO\_EXE}\\ \hline
        \texttt{24} & \texttt{vssadmin.exe resize shadowstorage /for=c: /on=c: /maxsize=401MB}\\ 
         \bottomrule
    \end{tabular}
    }
\end{table}

\end{document}